\documentclass[utf8]{frontiersFPHY} 

\usepackage{url,hyperref,lineno,microtype,subcaption}
\usepackage{xcolor}
\usepackage{color}
\usepackage{caption}
\usepackage{subcaption}
\usepackage{float}
\usepackage{graphicx}
\usepackage{amsmath,amssymb,epsfig,xspace,alltt,verbatim,bm}

\graphicspath{{APT_pics/}}

\definecolor{Mygreen}{rgb}{0.4,0.91,0.09001}

\newcommand{\PT}{\mathcal{PT}}
\newcommand{\p}{\mathcal{P}}
\newcommand{\T}{\mathcal{T}}
\newcommand{\tg}{\tilde{g}}

\def\keyFont{\fontsize{8}{11}\helveticabold }
\def\firstAuthorLast{Rodrigues {et~al.}} 
\def\Authors{A.S. Rodrigues\,$^{1,*}$, R. M.\ Ross$^{2}$, V.V.\
  Konotop $^3$, A.\ Saxena\,$^{4}$ and P.G.\ Kevrekidis\,$^{2}$}
\def\Address{$^{1}$Departamento de F\'{\i}sica e Astronomia/CF-UM-UP-CFP, Faculdade de Ci\^{e}ncias, Universidade do Porto, R. Campo Alegre,
	s/n - 4169-007 Porto, Portugal \\
$^{2}$Department of Mathematics and Statistics, University of Massachusetts,
	Amherst MA 01003-4515,  USA\\
$^3$Departamento de F\'{i}sica and Centro de F\'{i}sica Te\'orica e Computacional, Faculdade de Ci\^encias, Universidade de Lisboa, Campo Grande, Ed. C8, Lisboa 1749-016, Portugal\\
$4$Theoretical Division and Center for Nonlinear Studies, Los Alamos National Laboratory, Los Alamos, New Mexico 87545
  }
\def\corrAuthor{A.S. Rodrigues}

\def\corrEmail{asrodrig@fc.up.pt}

\begin{document}
\onecolumn
\firstpage{1}

\title[Nonlinear Anti-(Parity-Time) symmetric dimer]{Nonlinear Anti-(Parity-Time) symmetric dimer} 

\author[\firstAuthorLast ]{\Authors} 
\address{} 
\correspondance{} 

\extraAuth{}

\maketitle

\begin{abstract}

 

In the present work we propose a nonlinear anti-$\mathcal{PT}$-symmetric 
dimer, that at the linear level has been experimentally created in the
realm of electric circuit resonators.  We find four families of
solutions, the so-called upper and lower branches, both in a symmetric
and in an asymmetric (symmetry-broken) form. We unveil analytically
and confirm numerically the critical thresholds for the existence of
such branches and
explore the bifurcations (such as saddle-node ones) that delimit their
existence,
as well as transcritical ones that lead to their potential exchange of
stability. We find that out of the
four relevant branches,
only one, the upper symmetric branch, corresponds to a  spectrally and
dynamically
robust solution. We subsequently leverage detailed 
direct numerical computations in order to explore the dynamics of the
different states, corroborating our spectral analysis results.

\tiny
 \keyFont{ \section{Keywords:} anti-parity-time symmetry, nonlinearity, dimer, stability, symmetry breaking} 
\end{abstract}

\section{Introduction}
\label{sec:intro}

Dissipative systems, whose linear Hamiltonians obey
parity-time ($\PT$) symmetry are known to share properties of
Hermitian systems; indeed that was a central original motivation for the
proposal of such systems in connection to the foundations of
quantum mechanics~\cite{Bender1,Bender2}. More recently,
this proposal found a fertile ground for experimental realization in a diverse
array of other fields, including in optical media
(where loss and controllable gain are
ubiquitous)~\cite{Ruter,Peng2014,peng2014b,ncomms2015},
electronic circuits~\cite{Schindler1,Schindler2,Factor} and even
mechanical systems~\cite{Bender3}. A key feature that most of the
above systems share is the possibility to straightforwardly
include nonlinearity in the dynamics; e.g., in optical media, this can
be achieved via increase of the optical intensity.
This rendered the study of these nonlinear systems and of their
nonlinear modes/waveforms a canonical next
step within such studies.

Nonlinear dimers
(two-site-systems)~\cite{RaKot2010,lik,rodr,SuZhiKi2010,XuKeSax2015,BaPeDu2015}
and quadrimers~\cite{LiKeMaGu2012,KoZe2012,GuDeSa2015} are
among the simplest systems allowing one to observe the above mentioned
features. At this point in time, many of the relevant observations
have been summarized in comprehensive reviews~\cite{RevPT,KZY} and books~\cite{JY}. 

As is known from the above settings, 
specific symmetries of the underlying linear system impose constraints
on the existence as well as on the types of nonlinear modes sustained
by the system of interest. The literature mentioned above was mainly concerned with
parity ($\p$) - time ($\T$) symmetric systems, whose linear
Hamiltonians commute with the $\PT$-operator. In this work we address
the possibility of  anti-$\PT$ symmetry of the linear
Hamiltonian, as concerns the existence and stability of the associated
{\it nonlinear} modes. Such
dissipative systems in the linear setting were suggested
in~\cite{GeTur13}, and since then their experimental feasibility has
been argued in linear dissipatively coupled optical
systems~\cite{YaLiYo17} and illustrated in the context of a warm
atomic-vapour cell~\cite{PengXiao}. More recently,
they have been experimentally realized in a dimer of
resistively coupled amplifying RLC-circuits, where various intriguing
features such as corresponding exceptional points (EPs) and energy
difference conserving dynamics were identified~\cite{Choi18}. Further
attention to anti-$\PT$ symmetric systems was due to the possibility of their
usage for generating more sophisticated Hamiltonians, for example
odd-$\PT$-symmetric systems~\cite{KoZe2018,HaZeKoHuKo}, or for the simulation of
anti-parity-time symmetric Lorentz dynamics~\cite{Li2019}.
A relatively recent summary of the relevant activity can be found
in~\cite{lige}.

While a systematic effort has been made to explore anti-$\PT$
symmetric linear media, to the best of our knowledge, far less
of an effort has been invested in nonlinear analogues thereof.
It is toward that latter vein that our effort herein is geared. 
Specifically, we revisit the prototypical linear anti-$\PT$ model
motivated from the experimental realization of~\cite{Choi18}.
We endow the relevant model with nonlinearity which
is straightforward in the optical realm, as well as the atomic-vapour
setting~\cite{PengXiao},
but also genuinely feasible in the electrical circuit realm as
well; e.g., via the dependence of capacitances on the voltage
that has been used as a source of numerous nonlinear features in
such settings~\cite{rem}. For this nonlinear anti-$\PT$-symmetric
dimer, our aim is to explore the prototypical nonlinear states
thereof, as well as their spectral stability features and nonlinear
dynamical properties. The algebraic nature of the system
permits us to identify the associated nonlinear modes in an
exact analytical form. We indeed find two symmetric and two
asymmetric branches of solutions. Nevertheless, the corresponding
stability matrices cannot be diagonalized to yield the relevant
eigenvalues in a simple, explicit closed form. We thus
compute the relevant spectrum numerically. We find that out of the 
four branches of solutions {\it only one symmetric state is stable}.
Nevertheless, we also elucidate the complex bifurcation structure
of the model. Indeed, the two symmetric branches emerge through
a saddle-node (SN) bifurcation. The lower (unstable) symmetric branch
is also involved in a transcritical bifurcation with the asymmetric
branches, with the latter also terminating in a separate SN
bifurcation. 
We then go on to examine the dynamical evolution of both
stable and unstable states, corroborating the spectral results, but
also illustrating the fate of the unstable waveforms.

Our presentation is structured as follows. In Section 2, we briefly present
and explain the relevant mathematical model. In section 3, we analyze
the existence of its nonlinear solutions. In section 4, we again briefly discuss
the spectral linearization around such waveforms. In Section 5, we
present
our numerical stability and dynamical results. Finally, in section 6
we summarize our findings and present our conclusions as well
as some directions for future study.

\section{The model}

Bearing in mind optical applications to a two-waveguide
geometry~\cite{YaLiYo17}, atomic ones for a pair
of two collective spin-wave excitations \cite{PengXiao},
or a pair of RLC circuits per the experiment of~\cite{Choi18}, we
chose, arguably, the simplest model of an anti-$\PT$-symmetric dimer $\psi=(\psi_1,\psi_2)^T$ ($T$ stands for transpose) governed by the equation:
\begin{equation}
	i\frac{d \psi}{d z} = H_0\psi + F(\psi)\psi,  \qquad 
	H_0=  \begin{pmatrix} -\delta + i\gamma & iC \\ iC & \delta + i\gamma \end{pmatrix}
	\label{eq:eq_motion}
\end{equation}
where $C$, $\delta$ and $\gamma$ are real parameters describing
non-conservative coupling between the waveguides (or circuits),
difference of the propagation constants (it will be assumed without
loss of generality that $\delta>0$), and gain (if $\gamma>0$)  or loss
(if $\gamma<0$) in the waveguides (or circuits), respectively. {We notice, that while one of the parameters, say $\delta$, in $H_0$ can be scaled out, we keep all of them since they correspond to different physical processes, and thus facilitate interpretation of the results.} The
non-conservative 
nonlinearity in (\ref{eq:eq_motion}) is given by the diagonal matrix
\begin{equation}
	F(\psi) =  \begin{pmatrix} g|\psi_1|^2 + \tilde{g}|\psi_2|^2 & 0 \\ 0 & g|\psi_2|^2 + \tilde{g}|\psi_1|^2 \end{pmatrix}
\end{equation}
with $g = g_1 - ig_2$ and $g_2 > 0$ describes the nonlinear absorption ($g_{1,2}$ and  $\tilde{g}$ are real).
Defining the parity $\p=\sigma_3$ and time-reversal (anti-linear) operator $\T$ as a complex conjugation, $\T\psi=\psi^*$ one can verify that 
\begin{eqnarray}
	\PT H_0 +H_0\PT=0
\end{eqnarray}
We notice that the introduced system is characterized by the active
(non-Hermitian) coupling which was previously addressed in a number of
publications without~\cite{PengXiao,YaLiYo17,Choi18,Li2019} and with
conservative and non-conservative
nonlinear contributions~\cite{AlBaRaF2014}. It is relevant to mention in passing
that some of these works, including experimental ones such
as~\cite{PengXiao}
indicate how nonlinearity can be incorporated in the relevant
considerations
even though they do not study it in detail. Within the model
(\ref{eq:eq_motion}), nonlinearity stems from
self- and cross-phase modulation, characterized by the strengths $g_1$
and $\tg$, respectively, as well as from nonlinear absorption of strength $g_2$.

At the linear level, the eigenvalue problem for $H_0$  
\begin{align}
	H_0\varphi_j=b_j\varphi_j 
\end{align}
is readily solved
\begin{eqnarray}
 	b_\pm=i\gamma\pm\Lambda, \qquad \varphi_1=\left(\begin{array}{c}
		iC \\ \delta\pm\Lambda
	\end{array}\right)
\end{eqnarray}
where
\begin{align}
	\Lambda = \sqrt{C^2-\delta^2}
\end{align}
describes the deviation from the EP $\delta=|C|$ of the linear Hamiltonian.

\section{Nonlinear case steady state solutions}
\label{sec:nlinlim}

Turning to the nonlinear problem we start with steady state solutions of Eq.~(\ref{eq:eq_motion}) and employing the ansatz
\begin{equation}
	\psi = e^{ibz}\begin{pmatrix}\psi_{10}e^{i\phi/2} \\ \psi_{20}e^{-i\phi/2} \end{pmatrix},
	\label{eq:ansatz}
\end{equation}
where $b$ is a real spectral parameter and $\psi_{10}$ and $\psi_{20}$ are real, we obtain the system 
\begin{align}
	0 & = \Big(b-\delta +i\gamma + g|\psi_{10}|^2 + \tilde{g}|\psi_{20}|^2\Big) \psi_{10} + i C\psi_{20}e^{-i\phi} 
		\label{eq:st_state1}
	\\
	0 & = \Big( b + \delta + i\gamma +g|\psi_{20}|^2 + \tilde{g}|\psi_{10}|^2\Big) \psi_{20} + i C\psi_{10}e^{i\phi} 
	\label{eq:st_state2}
\end{align}

\subsection{Equal amplitude solutions}

Let us search for solutions with $\psi_{10} = \pm\psi_{20}$.  Since Eq.~(\ref{eq:eq_motion}) is invariant under the transformation $\psi \mapsto -\psi$, we simplify the analysis
by restricting our attention to the case $\psi_{10} \ge 0$. Now the system (\ref{eq:st_state1})-(\ref{eq:st_state2}) is reduced to two decoupled equations
\begin{align}
0=	b\mp \delta +i\gamma + (g+ \tilde{g})\psi_{10}^2+ \tilde{g} \pm i Ce^{\mp i\phi} 
	\label{eq:st_decoup}
\end{align}
that are readily solved, giving two steady state solutions
\begin{align}
	\label{equal-ampl}
	\psi_{10} &= \psi_{20} = \sqrt{\frac{\gamma  \pm \Lambda}{g_2}},     \quad\cos\phi=\pm\frac{\Lambda}{C}\quad \sin\phi = \frac{\delta}{C},
	\\
	b &= - \frac{\big(g_1 + \tilde{g}\big)}{g_2}(\gamma  \pm \Lambda)
	\label{eq:sym_sol}
\end{align}
We also observe that the solution $\psi_{10} = -\psi_{20}$ is obtained from (\ref{equal-ampl}) by the $-\pi/2$ phase shift. 

Thus in total there are two (nontrivial) symmetric solutions, and they
exist (i.e., have real propagation constant $b$) only for $|C| >
\delta$ (recall that $\psi_{10}$ is real).
Whether just one
of them exists or both of them is controlled
by the relative size of $\gamma$ and $\Lambda$.
That is, assuming that $g_2>0$, the solution with the ($-$) sign in
Eq.~(\ref{equal-ampl})
necessitates that $\Lambda < \gamma$ in order to be real. 

Interestingly, at the EP of the linear Hamiltonian, $\Lambda=0$ or $C^2 = \delta^2$ , the two solutions in (\ref{equal-ampl}) coalesce at 
\begin{align}
	\label{bifurc}
	\psi_{10}= \psi_{20} = \sqrt{\frac{\gamma}{g_2}}, \quad\phi = \frac{\pi}{2}, \quad
	b & = -\frac{g_1 + \tilde{g}}{g_2} \gamma  
\end{align}
Thus the EP of the linear problem is also a point of a SN
bifurcation, leading to the emergence of the
two symmetric nonlinear modes. If $\gamma >0 $ then both bifurcating solutions exist
(as long as $|C| > \delta$) and are nontrivial. We will restrict our
considerations in what follows to this case, while a corresponding
algebraic analysis can similarly be carried out for $\gamma<0$.
It should also be noted that while the branch with the $+$ sign
in Eq.~(\ref{equal-ampl}) will exist for all values of $|C| >
\delta$,
the one with $-$ sign will only survive up to the critical
point of $C^2=\gamma^2 + \delta^2$.

\subsection{Unequal amplitude solutions}

We now search for solutions with unequal intensities which can be presented in the form
\begin{equation}
	\psi_{10} = A \cos\left(\frac{\xi}{2}\right), \qquad \psi_{20} = A \sin\left(\frac{\xi}{2}\right)
	\label{eq:ansatz_uneq}
\end{equation}
Observing that $\xi=\frac{\pi}{2},\frac{3\pi}{2}$ results in the equal-amplitude solutions considered above, we now consider the cases $0<\xi < \frac{\pi}{2}$ and $\frac{\pi}{2} < \xi < \frac{3\pi}{2}$ (recall that $\psi_{10}>0$).

Substituting (\ref{eq:ansatz_uneq}) in (\ref{eq:st_state1})-(\ref{eq:st_state2}), multiplying the first of the obtained equations by $\cos(\xi/2)$ and the second one by $\sin(\xi/2)$, we get
\begin{align*}
	(-b+\delta)A\cos^2(\frac{\xi}{2}) & = i\bigg(\gamma A \cos^2(\frac{\xi}{2}) + C A e^{-i\phi}\sin(\frac{\xi}{2})\cos(\frac{\xi}{2})\bigg) +
	\bigg(g\cos^2(\frac{\xi}{2}) + \tilde{g}\sin^2(\frac{\xi}{2})  \bigg) A^3 \cos^2(\frac{\xi}{2}) \\
	(-b-\delta)A\sin^2(\frac{\xi}{2}) & = i\bigg(\gamma A \sin^2(\frac{\xi}{2}) + C A e^{i\phi}\cos(\frac{\xi}{2})\sin(\frac{\xi}{2})\bigg) +
	\bigg(g\sin^2(\frac{\xi}{2}) + \tilde{g}\cos^2(\frac{\xi}{2})  \bigg) A^3 \sin^2(\frac{\xi}{2})
\end{align*}

Adding the two equations, simplifying, and equating real and imaginary parts we obtain the equations:
\begin{align}
	-b &+ \delta \cos(\xi)  =  \Big[g_1 \bigg(1-\frac{1}{2}\sin^2(\xi)\bigg) + \frac{1}{2} \tilde{g} \sin^2(\xi)\Big]A^2  \label{a1}\\
	0 & = \gamma +  C\sin(\xi)\cos(\phi) - g_2 \bigg(1-\frac{1}{2}\sin^2(\xi)\bigg)A^2 \label{a2}
\end{align}

If instead we now subtract the two equations, again after
simplification, and equating real and imaginary parts we obtain this time:
\begin{align}
	\delta - b \cos(\xi) & =  C \sin(\xi)\sin(\phi) + g_1 \cos(\xi)A^2 \label{a3} \\
	 A^2 &= \frac{\gamma}{g_2}  
	 \label{b4}
\end{align}
 Using the last result in (\ref{a2}), and dividing by $\sin\xi$ we obtain
\begin{equation}
	2C \cos(\phi) + \gamma \sin(\xi)= 0. \label{b2}
\end{equation}

Using (\ref{b4}) also in (\ref{a1}) after simplifying we obtain:
\begin{equation}
	b = \delta \cos (\xi) - \frac{\gamma g_1}{g_2} -\frac{\gamma(\tilde{g}-g_1)}{2g_2} \sin^2 (\xi). \label{b1}
\end{equation}

Finally, using this value for $b$ on the left hand side (LHS) of (\ref{a3}) (as well as (\ref{b4})), canceling terms, and multiplying through by $2g_2/\sin(\xi)$ we obtain:
\begin{equation}
	2g_2\delta \sin (\xi) + \gamma (\tilde{g}-g_1)\sin (\xi) \cos (\xi) = 2 g_2 C \sin (\phi) \label{b3}
\end{equation}

So, by solving (\ref{b2}) and (\ref{b3}) we compute $\xi$ and $\phi$,
which can then be replaced in (\ref{b1}) to obtain $b$. Together with
(\ref{b4}) it gives the full solution
for the asymmetric waveforms (i.e., specifying $(A,b,\xi,\phi)$).

We can formally solve equation (\ref{b2}) to obtain:
\begin{equation}
	\sin \xi = -\frac{2C}{\gamma} \cos \phi, \qquad \cos \xi = \pm \sqrt{1 - \left(\frac{2C}{\gamma}\right)^2 \cos^2 \phi}. 
	\label{eq:cosines}
\end{equation}

Then, inserting this result into (\ref{b3}), and rearranging we obtain:
\begin{equation}
  \sin \phi + \Delta \cos \phi \pm G\cos \phi \sqrt{1-\bar{C}^2 \cos^{2} \phi} = 0 
  \label{eq:newton1}
 	 \end{equation}
  where we defined $G\equiv (\tilde{g}-g_1)/g_2$, $\Delta \equiv \delta/(\gamma/2)$, and $\bar{C} \equiv 2C/\gamma$.  

We recognize
from the two signs in the above algebraic equations (resulting from,
e.g., Eq.~(\ref{eq:cosines}))
that two asymmetric solution families can be obtained from the above
analysis.  We now proceed to set up and subsequently explore the
stability of these four (two symmetric and two asymmetric) families of
solutions.

\section{Stability matrix}

The solutions found above need to be analyzed for their stability, in
order to assess their potential dynamical robustness. This is achieved by studying the eigenvalues of the  stability matrix, given by $\bar{\bar{M}} = \nabla_{\vec{u}} \vec{F}(\vec{u})$, where $\vec{u}$ is a steady state solution in the form $\vec{u}=(\psi_1,\psi_2,\psi_1^{*},\psi_2^{*})^T$, and $\vec{F}=(F_{\psi_1},F_{\psi_2},F_{\psi_1^{*}},F_{\psi_2^{*}})=(F_{\psi_1},F_{\psi_2},-F^{*}_{\psi_1},-F^{*}_{\psi_2})$.

For the anti-$\PT$ symmetric equations this has the form:
\begin{equation*}
	\bar{\bar{M}} =
	\begin{bmatrix}
		\bar{\bar{A}}  &  \bar{\bar{B}} \\
		-\bar{\bar{B}}^{*}  &  -\bar{\bar{A}}^{*} 
	\end{bmatrix}
\end{equation*}
where:
\begin{equation*}
	\bar{\bar{A}} =
	\begin{bmatrix}
		-\delta + i\gamma +b +2g|\psi_1|^2 +\tilde{g}|\psi_2|^2  &  \tilde{g}\psi_1 \psi_2^{*}  + iC \\[6mm]
		\tilde{g}\psi_1^{*} \psi_2 + iC &  \delta + i\gamma +b + 2g|\psi_2|^2 +\tilde{g}|\psi_1|^2 
	\end{bmatrix}
\end{equation*}
and
\begin{equation*}
	\bar{\bar{B}} =
	\begin{bmatrix}
		g\psi_1^2 & \tilde{g}\psi_1 \psi_2 \\[4mm]
		\tilde{g}\psi_1 \psi_2 & g\psi_2^2 .
	\end{bmatrix}
\end{equation*}

Using the ansatz $\vec{u} = \hat{e}_{\lambda} e^{\lambda z}$ in the equation for the stability  
\begin{equation*}
	i \frac{d\vec{u}}{d z} =  \nabla_{\vec{u}} \vec{F}(\vec{u}) \vec{u}, 
\end{equation*}
we obtain the eigenvalue problem as follows
\begin{equation*}
	\lambda \hat{e}_{\lambda}  = (-i \bar{\bar{M}}) \hat{e}_{\lambda} \equiv \bar{\bar{M}}' \hat{e}_{\lambda}.
\end{equation*}

Thus the eigenvalues of $\bar{\bar{M}}$ are the eigenfrequencies of the problem ($\omega$), while those of $\bar{\bar{M}}'= -i \bar{\bar{M}}$ are its eigenvalues ($\lambda$). 
[The two are connected via $\lambda=-i \omega$]. 

\section{Numerical results}
\label{sec:numerics_nn}

We look for solutions of the asymmetric form by performing a Newton search
of the algebraic equations~(\ref{eq:newton1}), followed by
continuation in the parameter $C$ of any solution thus found. For the
symmetric solutions, we did the same, although we could simply use our explicit analytical expressions
within the stability matrix (in order to identify their spectral
stability properties).

Below we present some representative results for the parameter values $\gamma=1.0$, $\delta=0.1$, $g_1 = 0.3$, $g_2 = 0.4$ and $\tilde{g}= 0.5$. $C$ is scanned from
$\delta$ up to $\sqrt{\delta^2 + \gamma^2}$, which are the limits for
a real amplitude for the ``negative" branch of the symmetric solution,
as can be seen in Eq.~(\ref{eq:sym_sol}).
Given our analytical formulae and numerical setup, similar findings
can be obtained for other parameter values.

\begin{figure}[h]
	\centering
		\centering
		\includegraphics[scale=0.32]{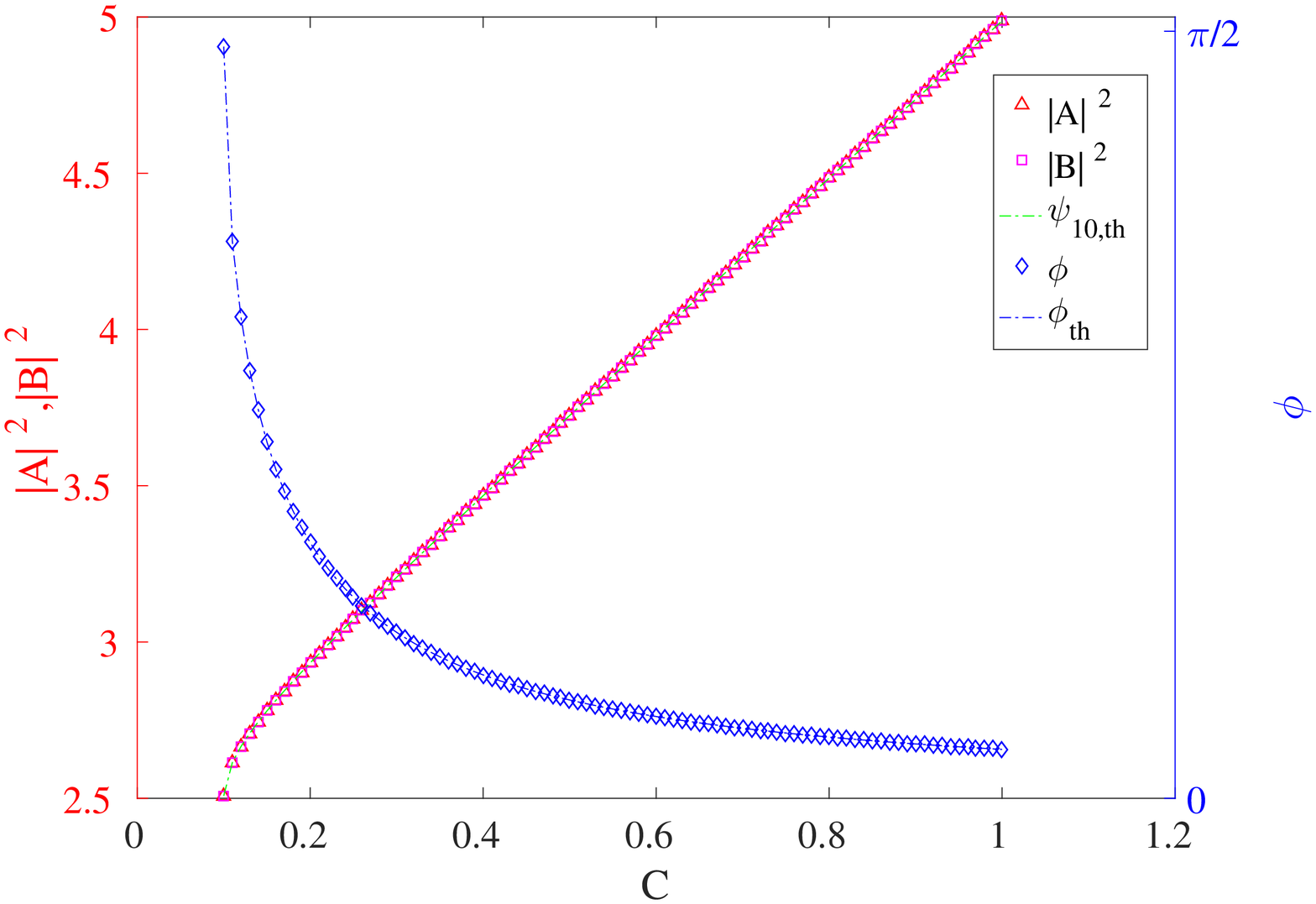}
		\centering
		\includegraphics[scale=0.32]{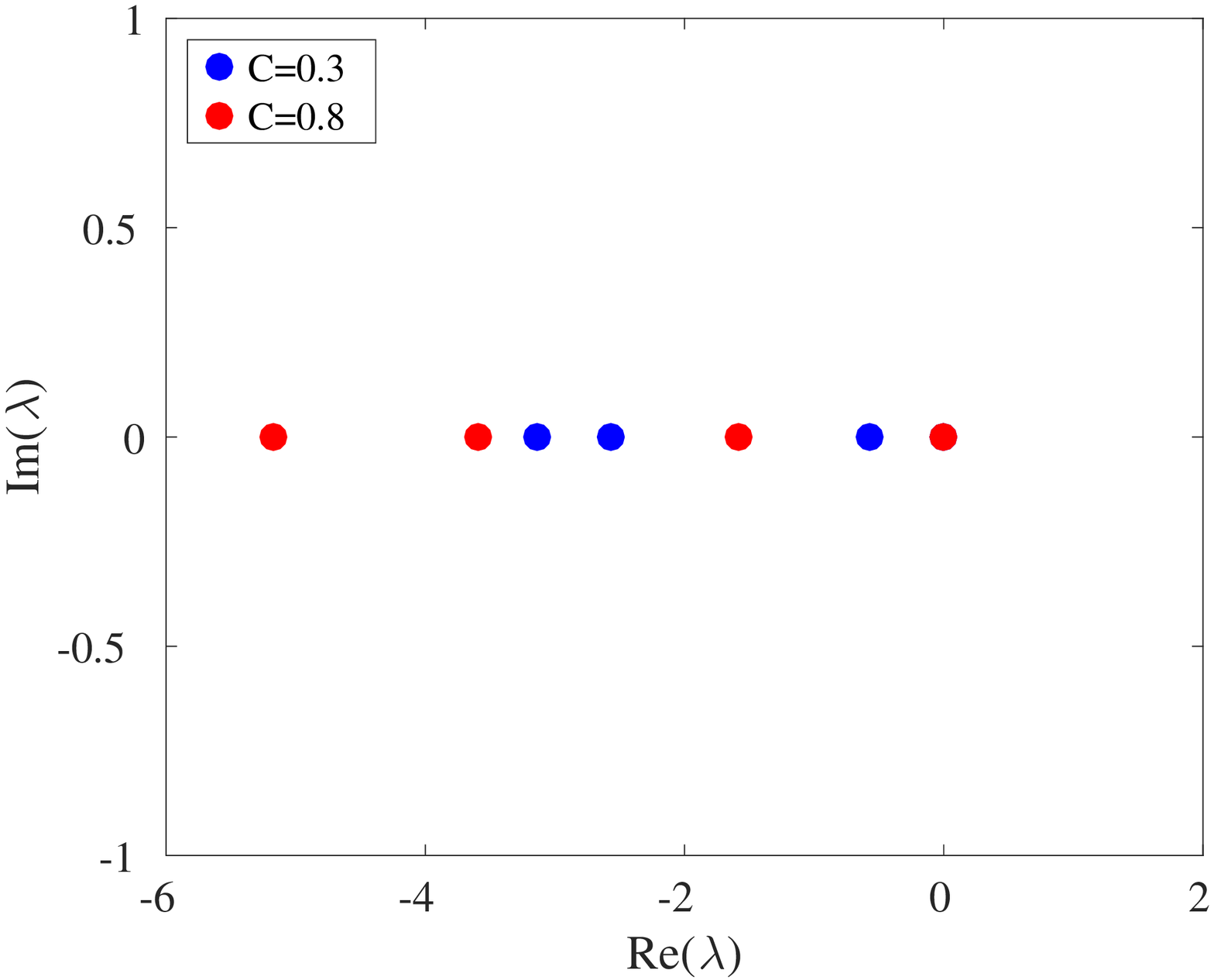}
	\centering
	\caption{(a) Amplitude and phase of steady state symmetric
          solutions from the upper symmetric branch. (b) Spectral
          plane
          (Re$(\lambda)$,Im$(\lambda)$) representation of the
          eigenvalues $\lambda$ for $C=0.3$ and $C=0.8$. Other parameter values are $\gamma=1.0$, $\delta=0.1$, $g_1 = 0.3$, $g_2 = 0.4$ and $\tilde{g}= 0.5$.}
	\label{fig:symua} 
\end{figure}

\begin{figure}[h]
	\centering
	\centering
		\centering
		\includegraphics[scale=0.35]{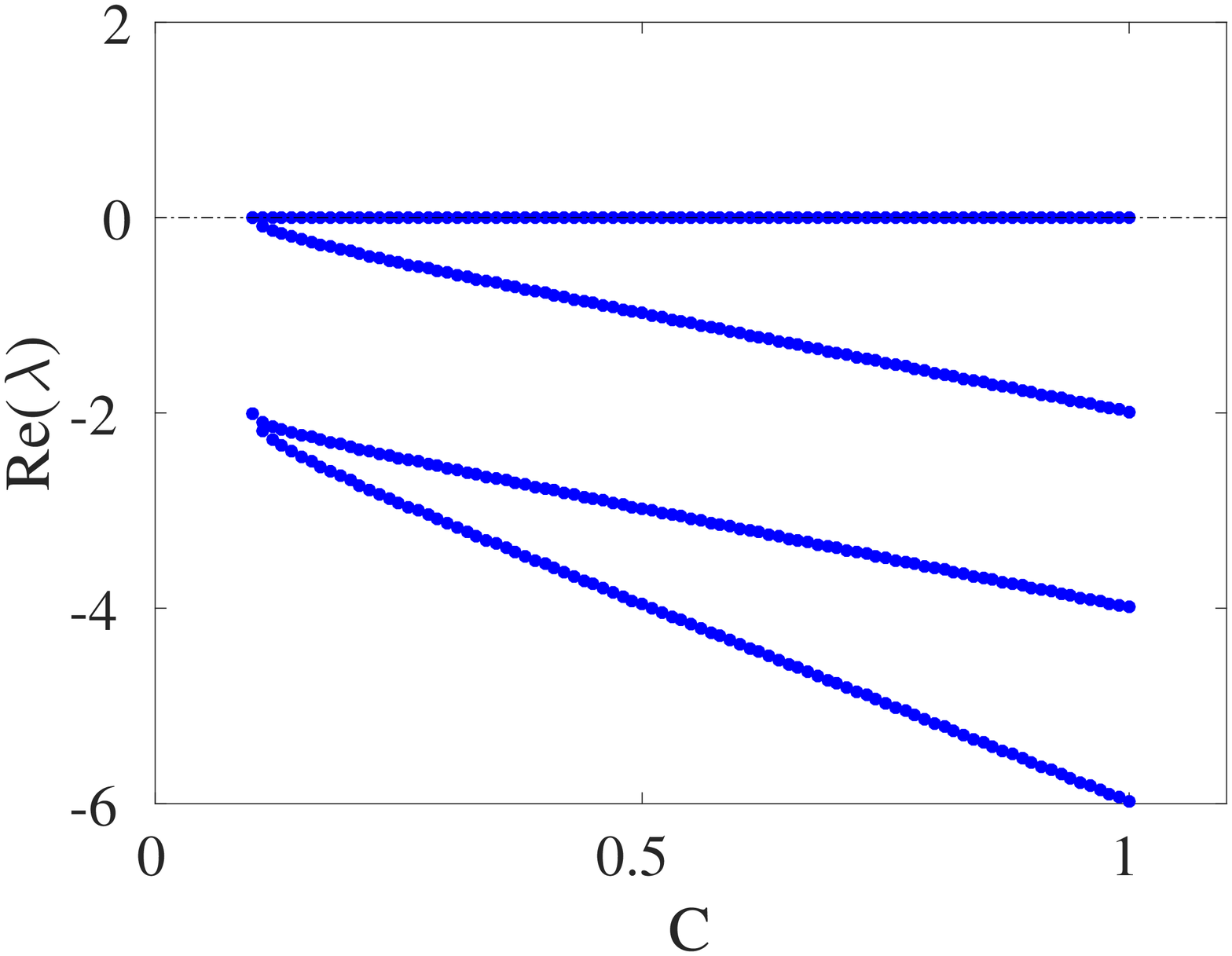}
		\centering
		\includegraphics[scale=0.35]{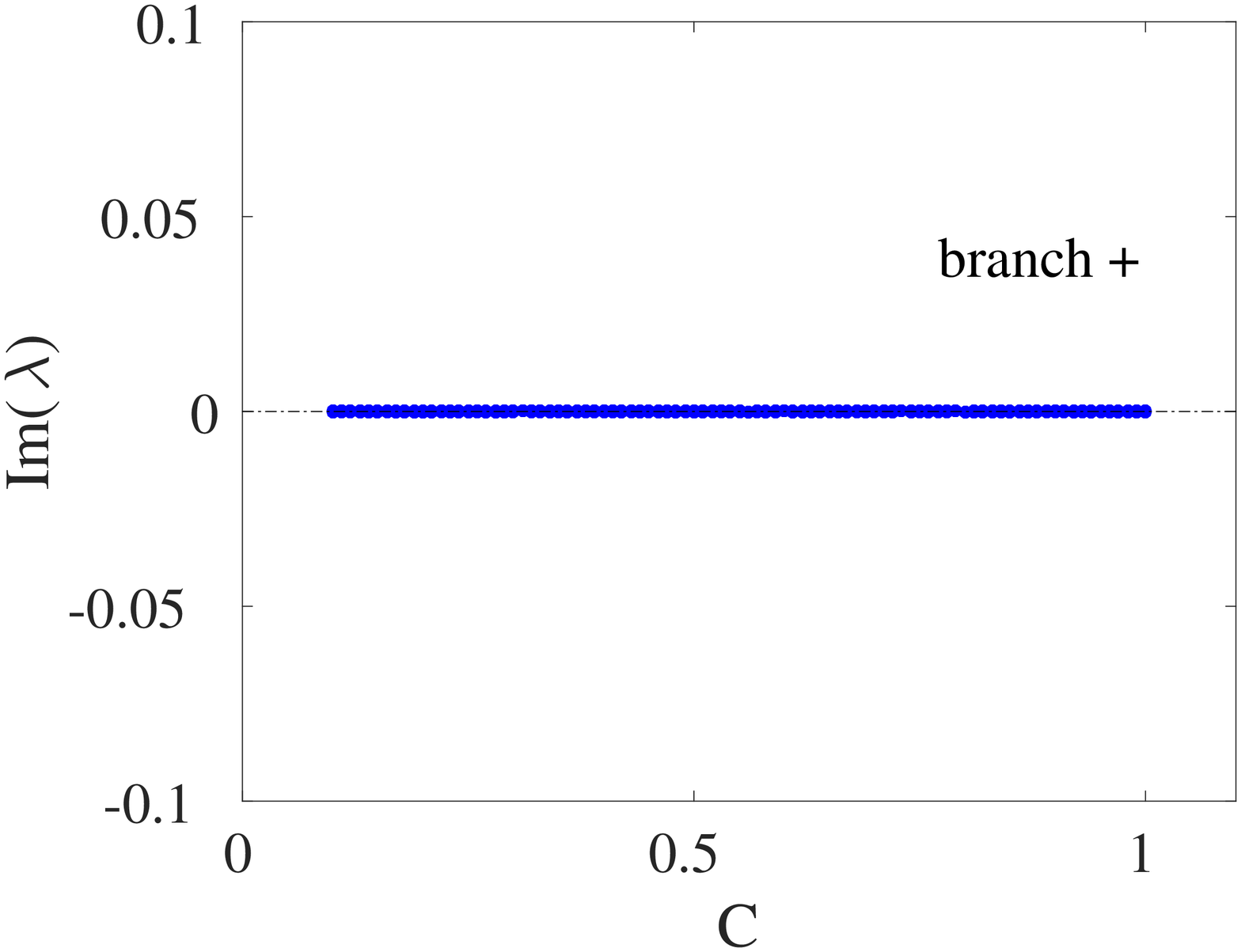}
	\centering
	\caption{Dependence of the real and imaginary parts of the
          eigenvalues of steady state symmetric solutions from the upper
          $(+)$ branch.
          The  other parameter values remain as in the caption of
          Fig.~\ref{fig:symua}.} 
	\label{fig:symub}
\end{figure}

For the symmetric case,
Figs.~\ref{fig:symua} and~\ref{fig:symub} show the results for the
``positive" branch (hereafter termed ``upper''). Represented are the amplitudes of the two nodes ($|A|^2$ and $|B|^2$), the phase difference between them ($\phi$) (left panel), and the complex plane representation of the eigenvalues for two values of the scanned parameter, $C$ (right panel). Then, in the second figure we show the dependence on $C$ of the real and imaginary parts of the eigenvalues.
One can observe that the amplitude grows with $C$, while the phase
difference (right axis) varies from $\pi/2$ to a little above
zero. Superposed to the numerical results are those of the analytic
expressions found above, and we can see that the match is very
good, as is of course expected.
The right panel of Fig.~\ref{fig:symua} shows that the eigenvalues are purely real
and indeed, as shown in Fig.~\ref{fig:symub}, they remain real throughout.

Fig.~\ref{fig:symub} shows the real and imaginary parts of the
eigenvalues as a function of $C$. Given that the largest value of the real part is zero, this upper branch
is spectrally stable. That is, all the relevant eigendirections are
associated with decay, aside from a neutral one (associated with
an overall phase freedom). Recall that this is a non-conservative
system, hence the relevant eigenvalues have to be in the left-half
of the spectral plane for stability, as is the case for this
branch. Indeed, we will see below that this is the only spectrally stable branch of this nonlinear anti-$\PT$-symmetric dimer.

\begin{figure}[H]
	\centering
		\centering
		\includegraphics[scale=0.32]{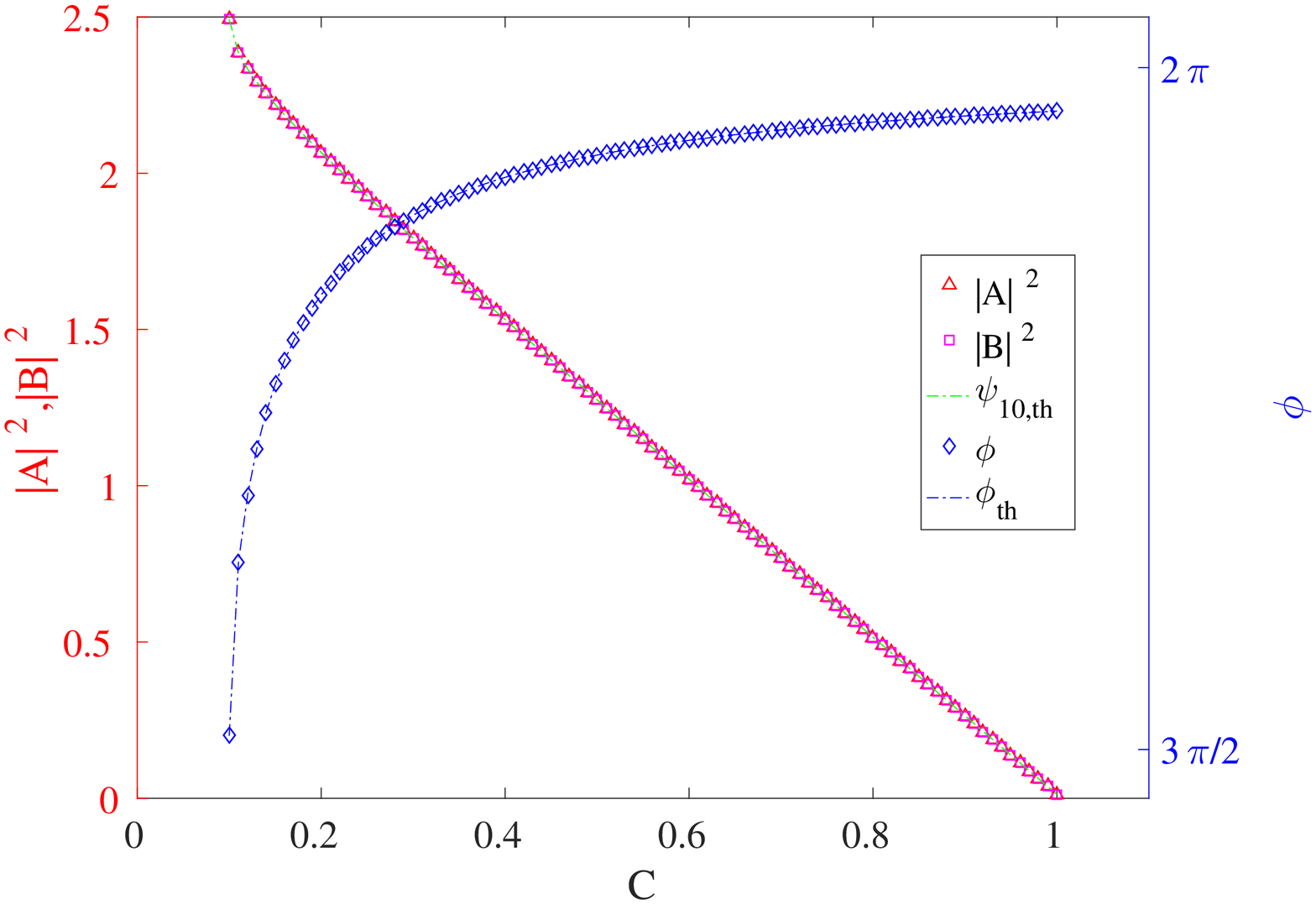}
		\includegraphics[scale=0.32]{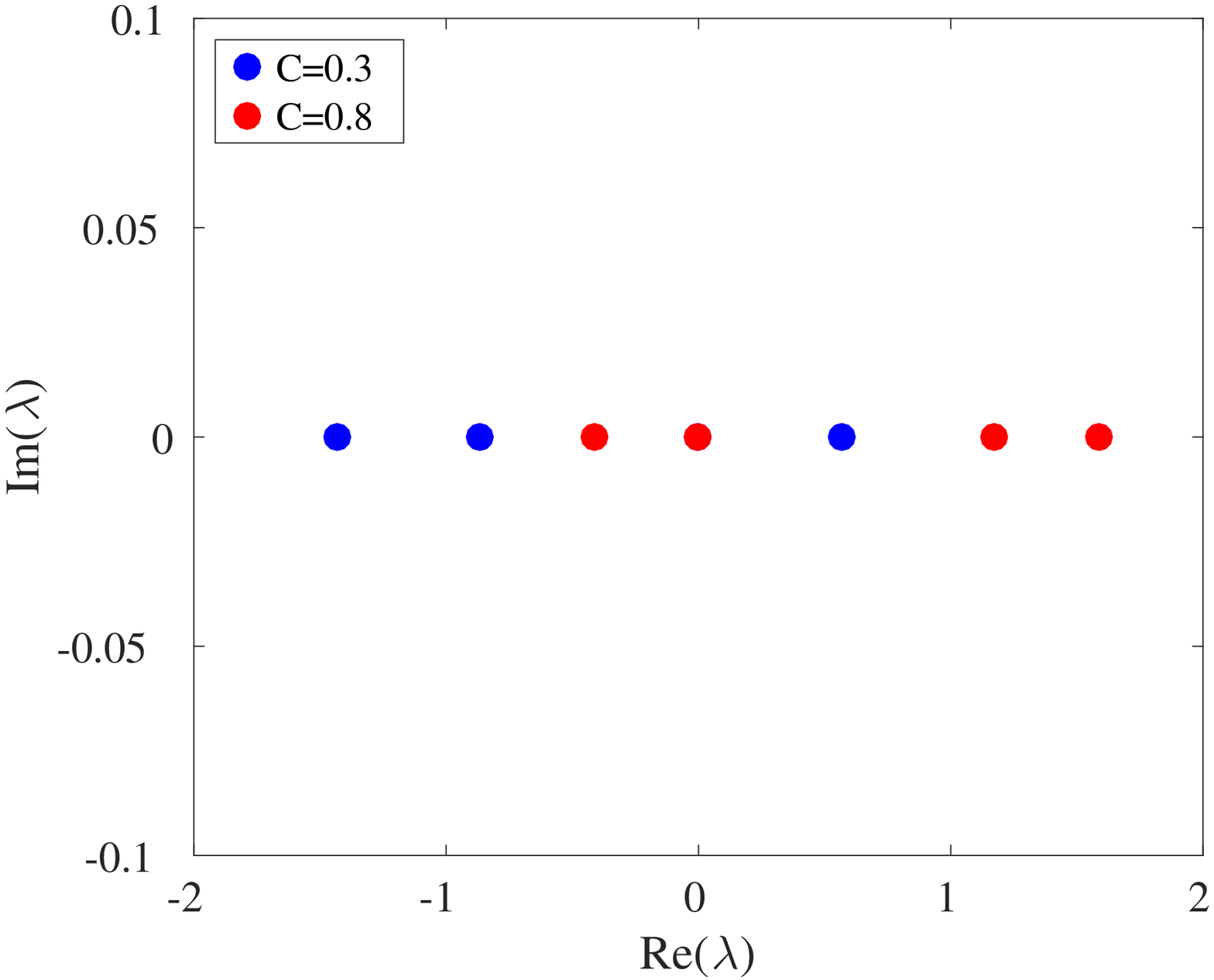}
	\caption{(a) Amplitude and phase of steady state symmetric
          solutions from
          the lower branch. (b) Spectral plane representation of
          eigenvalues for $C=0.3$ and $C=0.8$. See text for other parameter values. }
	\label{fig:symla}
\end{figure}

\begin{figure}[H]
	\centering
		\centering
		\centering
		\includegraphics[scale=0.35]{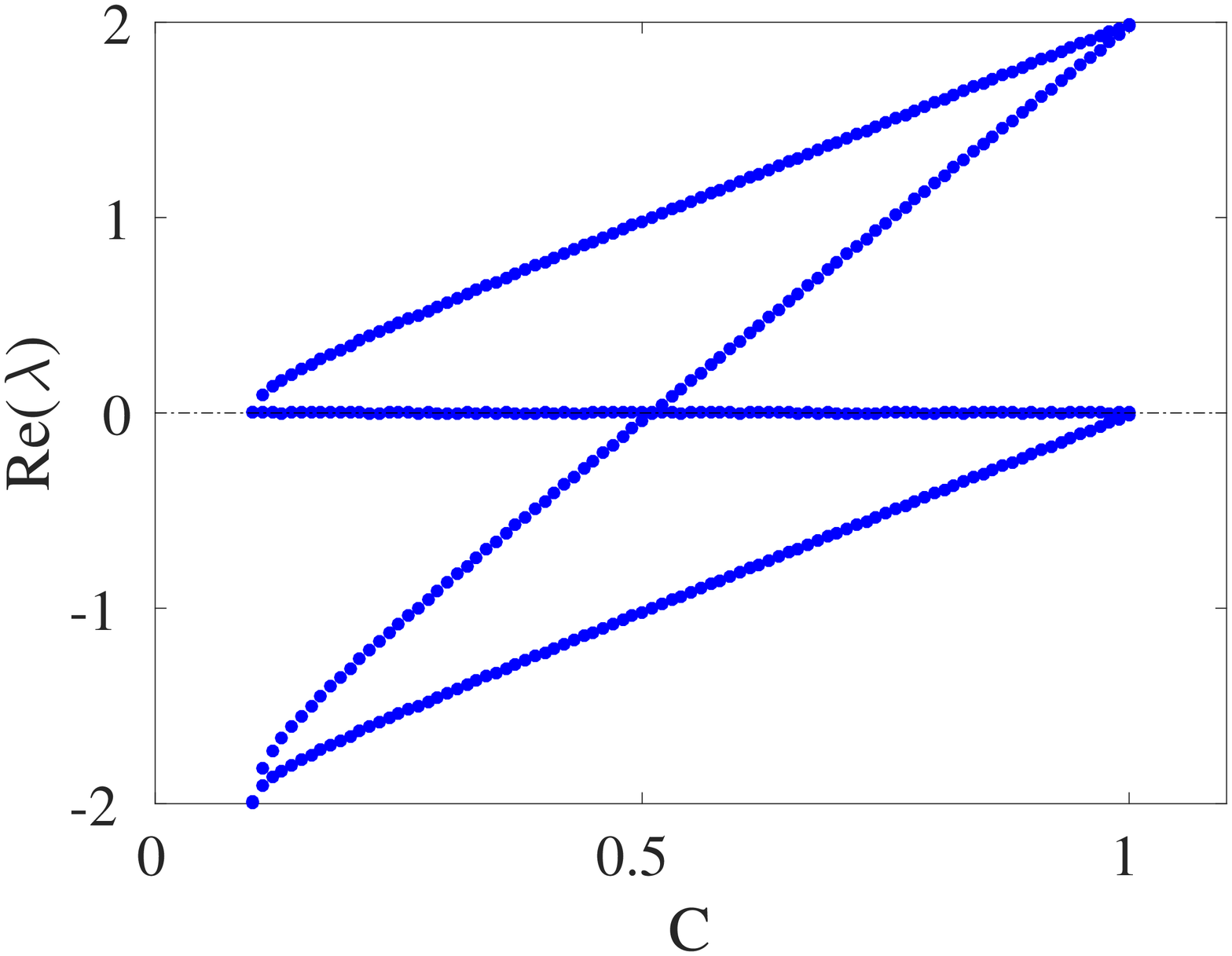}
		\centering
		\includegraphics[scale=0.35]{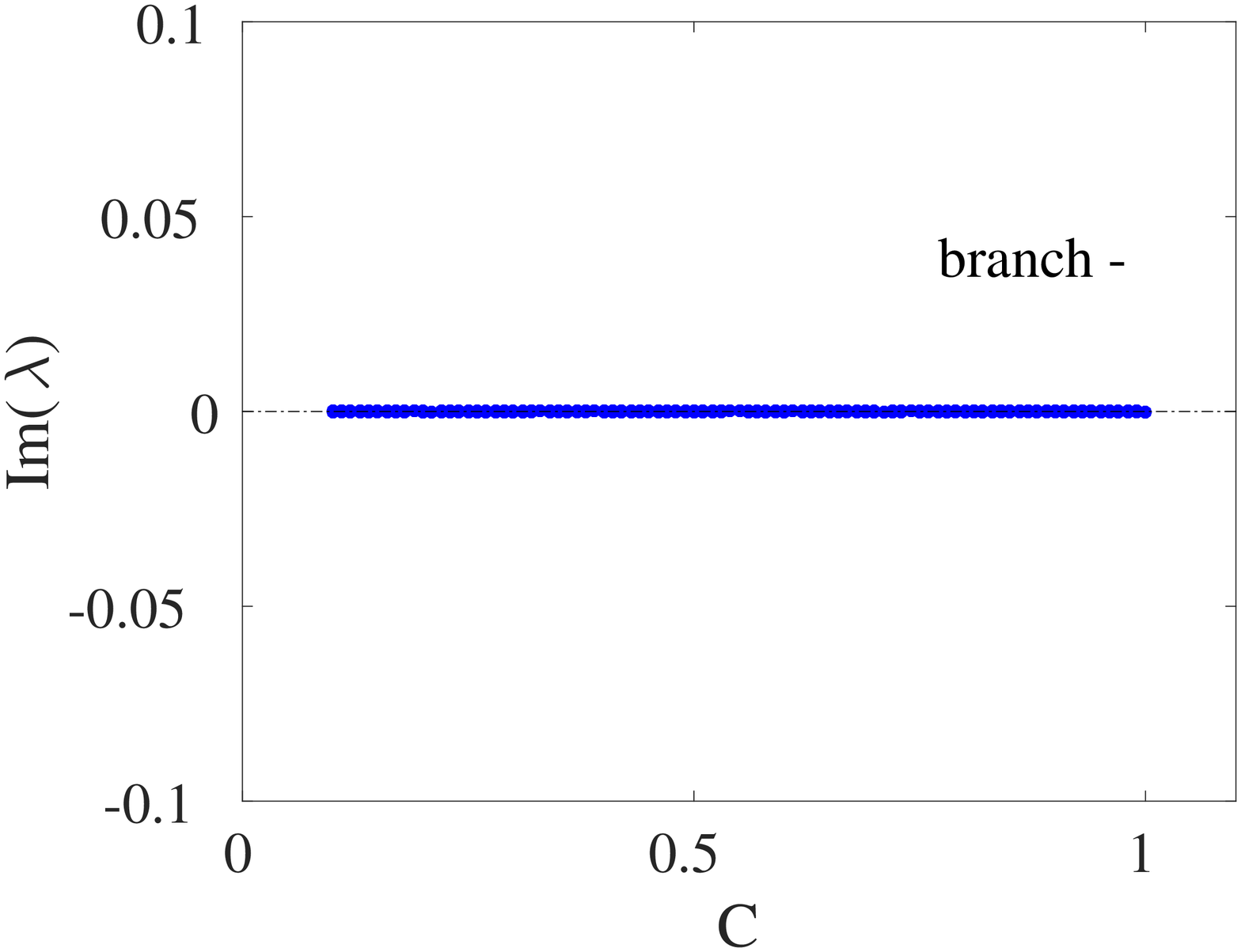}
	\centering
	\caption{Linear spectra of steady state symmetric solutions
          from the lower branch.
Importantly, in addition to the instability starting at
the saddle-node bifurcation which gives rise to its existence, the
solution
inherits an additional instability at a bifurcation point of $C = 0.51$. 
The rest of the parameters are the same as in the previous
Figures.} 
	\label{fig:symlb}
\end{figure}

In Figs.~\ref{fig:symla}-\ref{fig:symlb} we illustrate the
corresponding
results for the ``negative" solution in Eq.~(\ref{equal-ampl}), i.e., the
hereafter termed lower branch.
This time the amplitude decreases with increasing $C$, and the phase
difference increases from just over $\pi/2$ to $\pi$. The continuation
was started a little above $C=\delta$; for $C=\delta$ we would expect
both solutions to have $\phi=\pi/2.$
It is relevant to also note that the branches of Figs.~(\ref{fig:symua})-(\ref{fig:symla})
coincide at the critical point of $C=\delta$ at which the relevant
SN bifurcation arises with the upper branch corresponding to the node,
while the lower one to the saddle. In accordance with this picture
the spectra show again a purely real set of eigenvalues and in 
Fig.~\ref{fig:symlb} with one of them being positive and hence
corroborating the instability of the saddle $(-)$ symmetric
configuration of the lower branch. 

This is confirmed systematically also in Fig.~\ref{fig:symlb}, where
the
relevant unstable eigenvalue is seen to grow from $0$ beyond the
bifurcation point. Interestingly, an additional unstable
eigendirection
arises at some intermediate value of $C$ as well, rendering the
relevant branch more unstable.
We will return to the latter more elaborate bifurcation shortly.
Nevertheless, for the interval
of values of $C$ considered, the former instability is always
stronger (i.e., has a higher growth rate) than the latter one. 


\begin{figure}[!ht]
	\begin{minipage}{\textwidth}
		\centering
		\includegraphics[width=.4\textwidth]{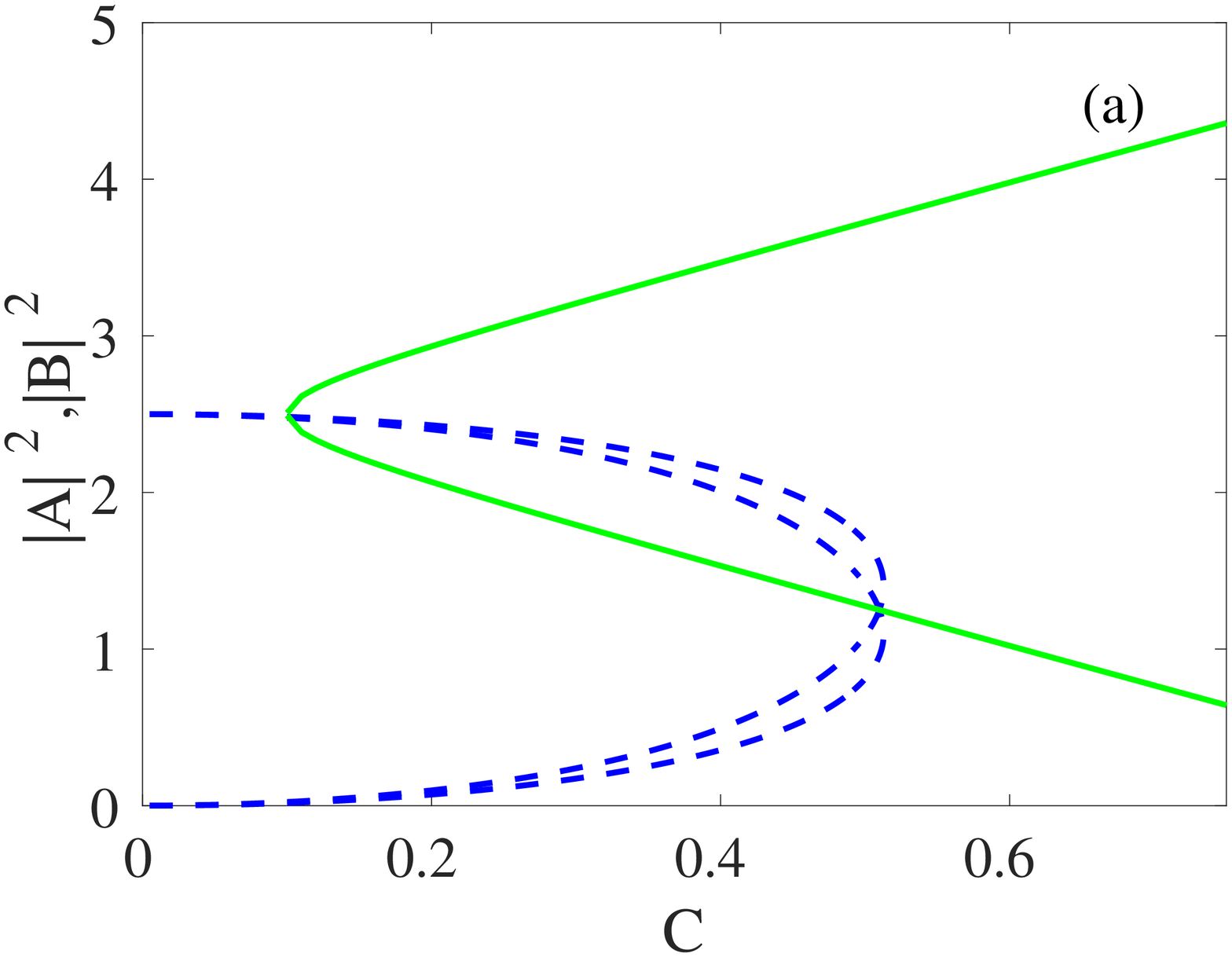}\quad
		\includegraphics[width=.41\textwidth]{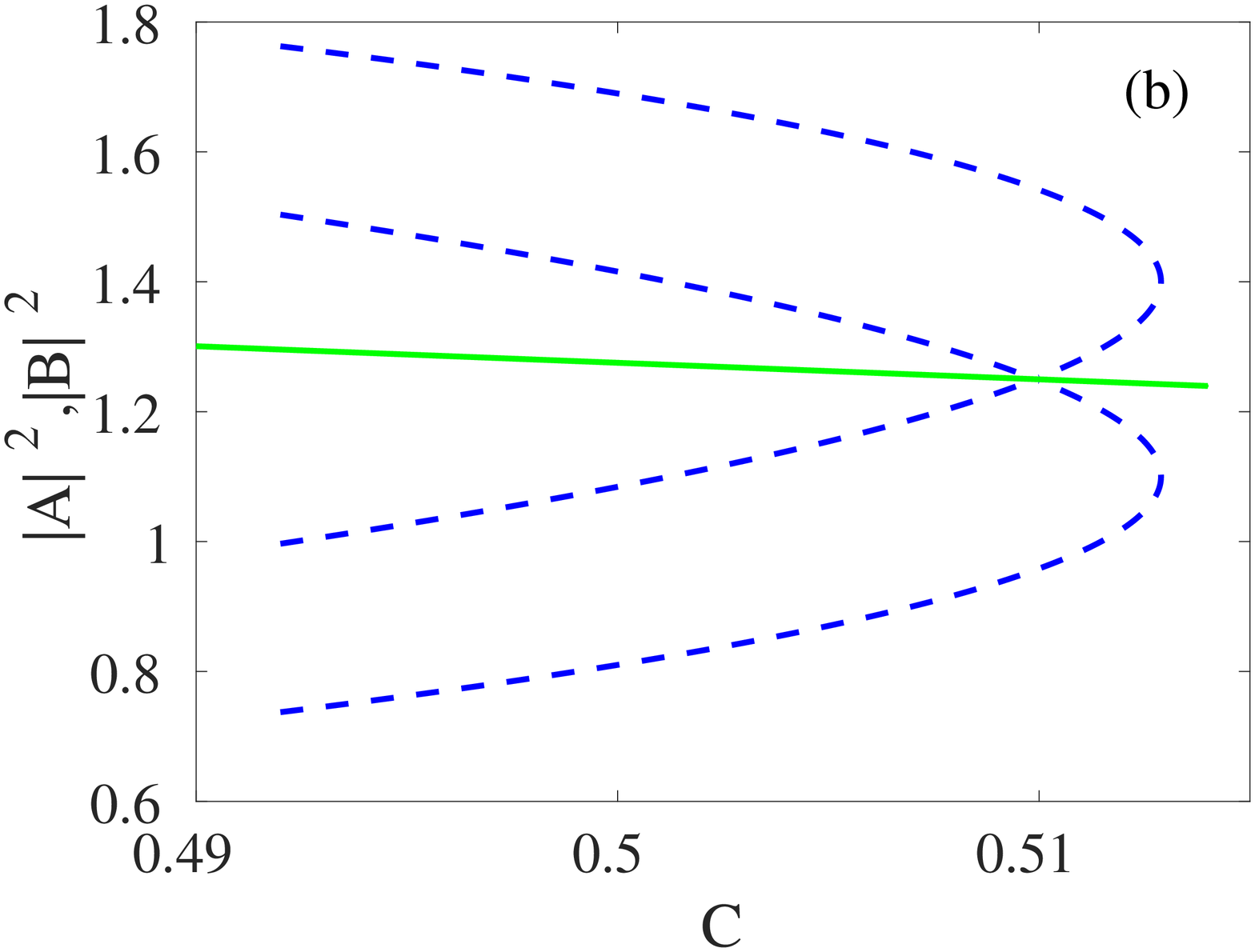}
		\caption{{Bifurcation diagram as  a function of (control)
					parameter $C$. (a) amplitudes
                                        $|A|^2,|B|^2$ ; (b) zoom in
                                        for the region where bifurcations
                                        occur. We do not show here the
                                        results
                                        for $\phi$ and $\xi$, although
                                        the same bifurcation features can be
                                        observed therein.}}
		\label{fig:bifurcC}
	\end{minipage}
\end{figure}

Now we turn to the results obtained for the two branches of
asymmetric solutions. Here, the bifurcation picture is far more
elaborate.
The bifurcation diagram as a function of the parameter $C$ is shown
  in Fig.~\ref{fig:bifurcC}.
  Let us note that in this diagram the symmetric (node upper and saddle
  lower) branches are also shown and their SN bifurcations are shown
  via the green (solid) curves, while the asymmetric branches are
  shown
  with blue (dashed) lines.
  The main feature of the latter is that there is a 3-way collision
  between the 2 asymmetric branches and the lower symmetric one, close
  to $C_{cr}^{(1)}=0.51$. This is easily seen in the detailed  (right) panel for
  the amplitude
  dependence.  The upper asymmetric branch goes past a fold en route
  to that collision, existing as a solution up to $C_{cr}^{(2)}=0.5128$.
  Indeed, the latter point is associated with a SN bifurcation
  corresponding to the termination (in terms of values of $C$) of
  the upper asymmetric branch. That is, the relevant branch does not
  exist for higher $C$ values. Interestingly, 
  the algebraic picture is somewhat more complicated in that 
  when solving Eq.~(\ref{eq:cosines}), the upper branch goes past
  the turning point of $C_{cr}^{(2)}$ and upon turning around collides
  with the lower branch $C_{cr}^{(1)}$. Nevertheless, this is, in a sense, an
  ``artifact'' of the closed form formulae of the analytical solutions
  of Eq.~(\ref{eq:cosines}). Observing the curves in the bifurcation
  diagram of Fig.~\ref{fig:bifurcC}, one can see that the ``inner''
  curves (the ones closer to the green line before $C_{cr}^{(1)}=0.51$
  at this critical point) collide between them and therefore
  become instantaneously symmetric before smoothly
  continuing en route to the collision with the ``outer'' (top and
  bottom) curves at $C_{cr}^{(2)}=0.5128$. That is to say, the
  former critical point signals a transcritical bifurcation,
  between the asymmetric and the symmetric branch, while the
  latter critical point signals a SN bifurcation leading to the
  termination of asymmetric branches.

\begin{figure}[h]
	\centering
		\centering
		\includegraphics[scale=0.41]{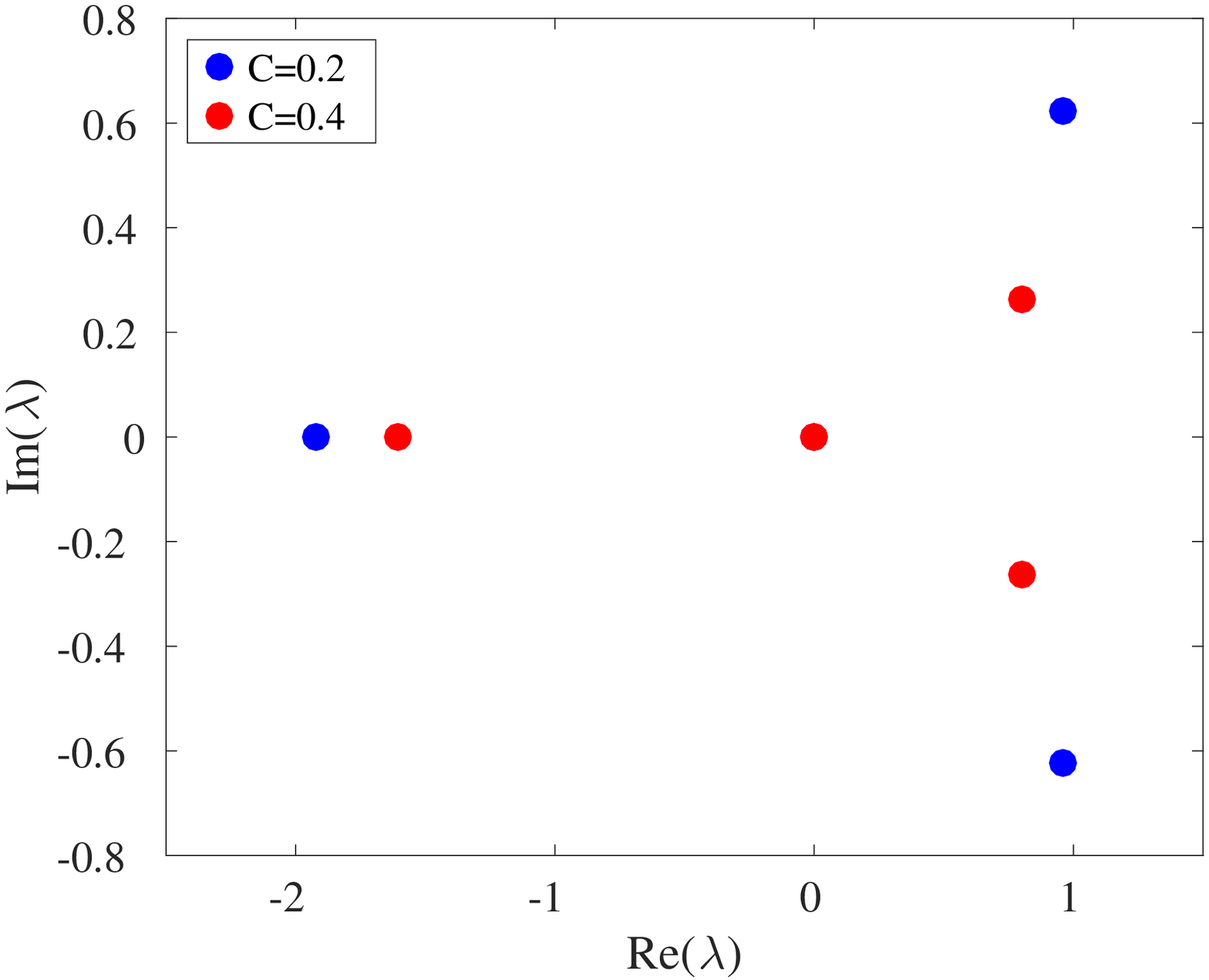}
	\includegraphics[scale=0.41]{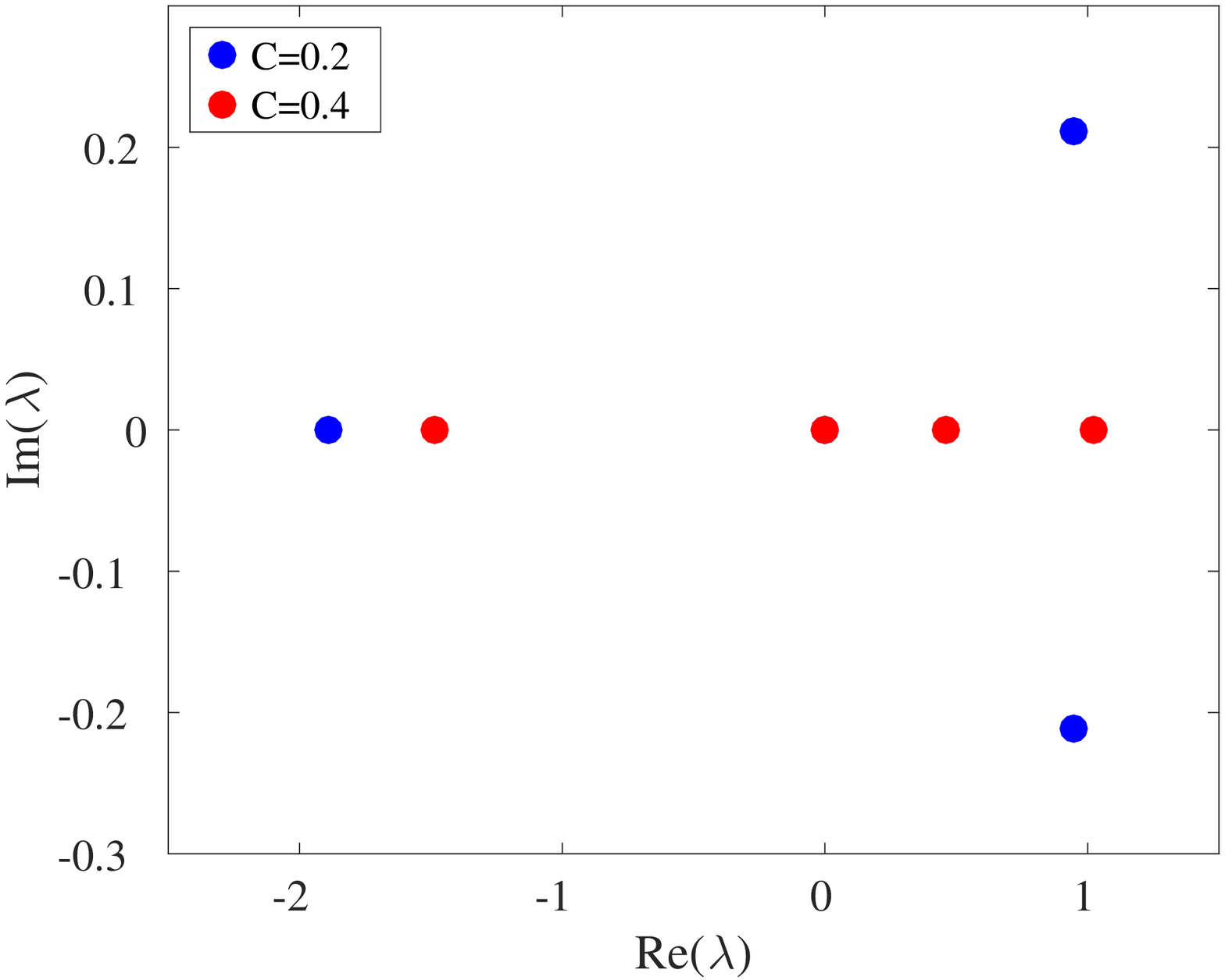}
	\caption{(a) and (b) Spectral plane representation
          of eigenvalues for $C=0.2$ and $C=0.4$ for the upper and
          lower asymmetric branches, respectively. The rest of the
          parameters
          are the same as in the previous figures.
}
	\label{fig:asymua}
\end{figure}

  Indeed, this picture is corroborated by the relevant eigenvalue
  plots.
  Figure~\ref{fig:asymua} illustrates some prototypical examples of
  the spectral plane of the upper and lower asymmetric solutions.
  Both of them bear a complex eigenvalue pair (i.e., are associated
  with an oscillatory instability featuring both growth and
  oscillation,
  as we will also see below). However, in the case of the upper
  branch this instability is persistent up to $C\approx 0.430$,
  while in the lower branch, it splits into two real eigenvalues
  earlier (parametrically),
  i.e., for $C\approx 0.287$. Notice, accordingly, the difference
  for the red points of $C=0.4$ in Fig.~\ref{fig:asymua}.

\begin{figure}[!ht]
	\begin{minipage}{\textwidth}
		\centering
		\includegraphics[width=.4\textwidth]{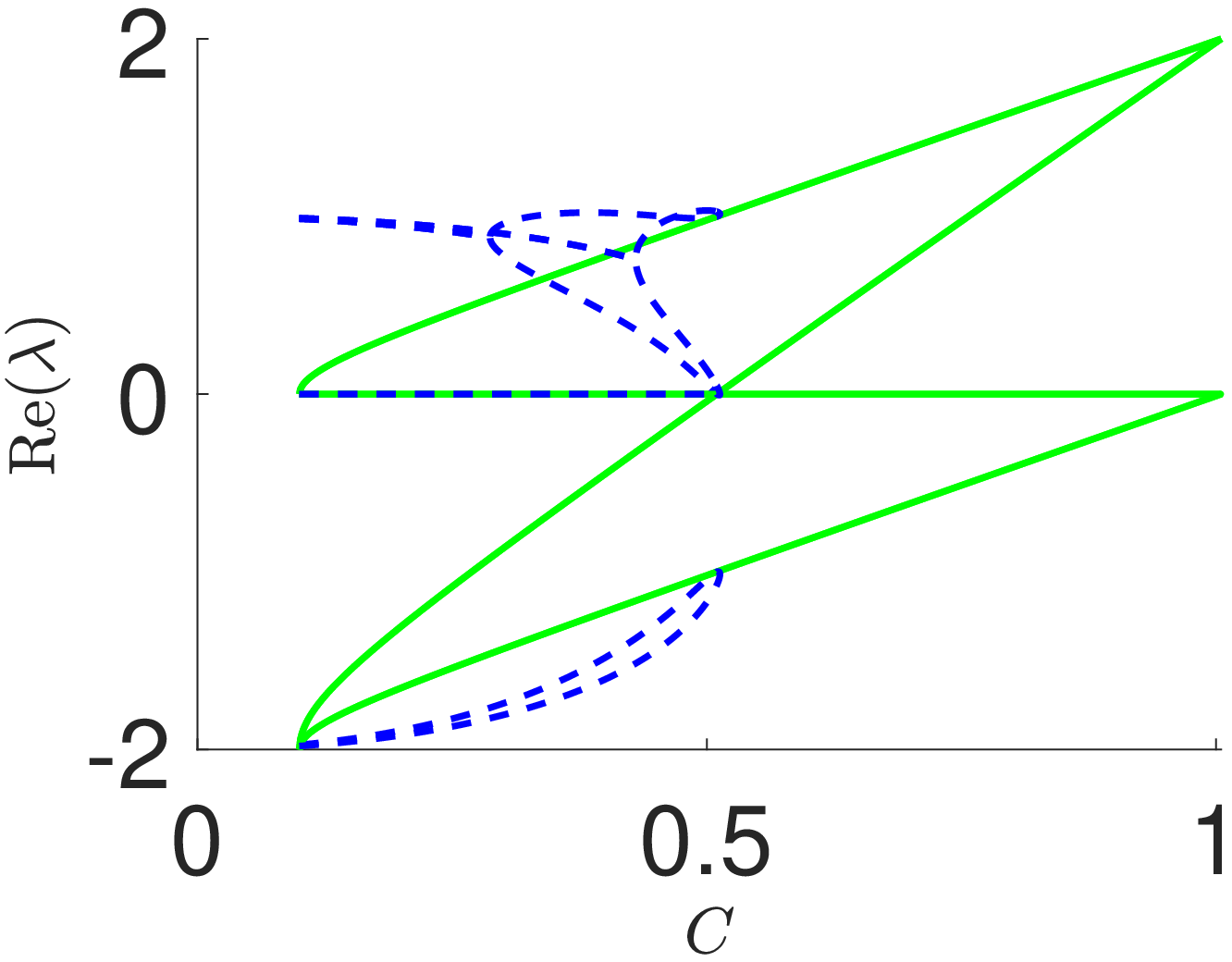}\quad
		\includegraphics[width=.4\textwidth]{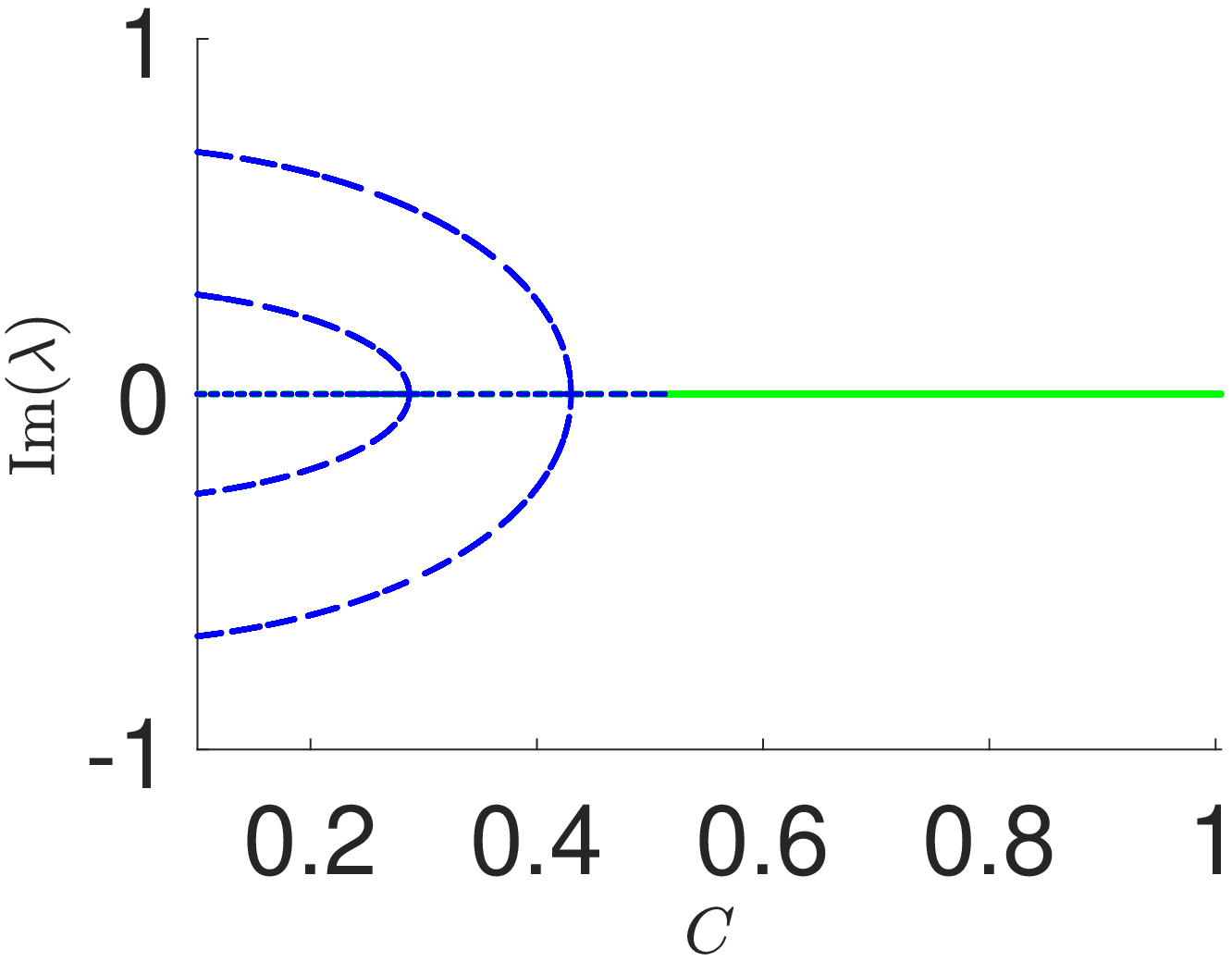}\\
                \includegraphics[width=.4\textwidth]{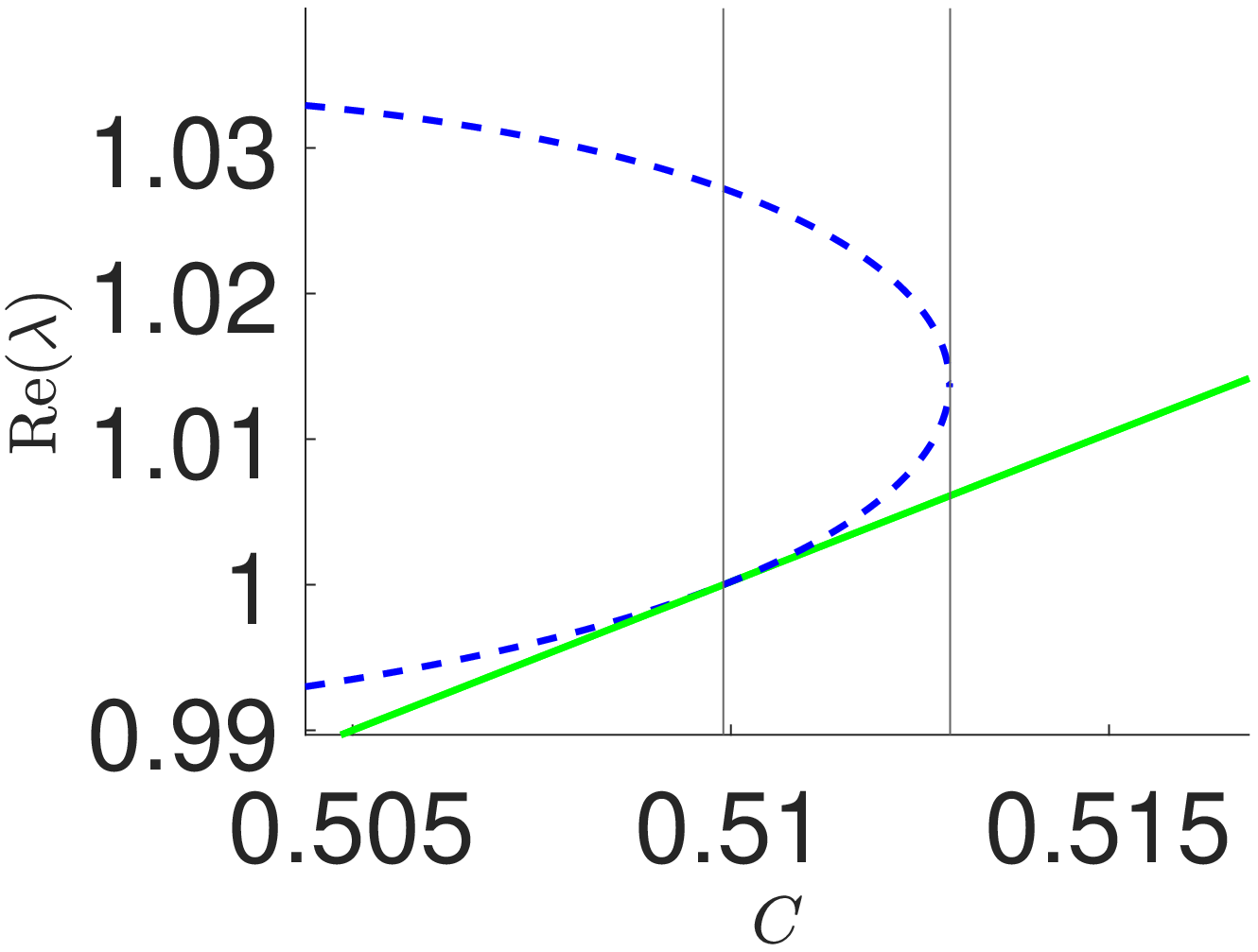}\quad
		\includegraphics[width=.4\textwidth]{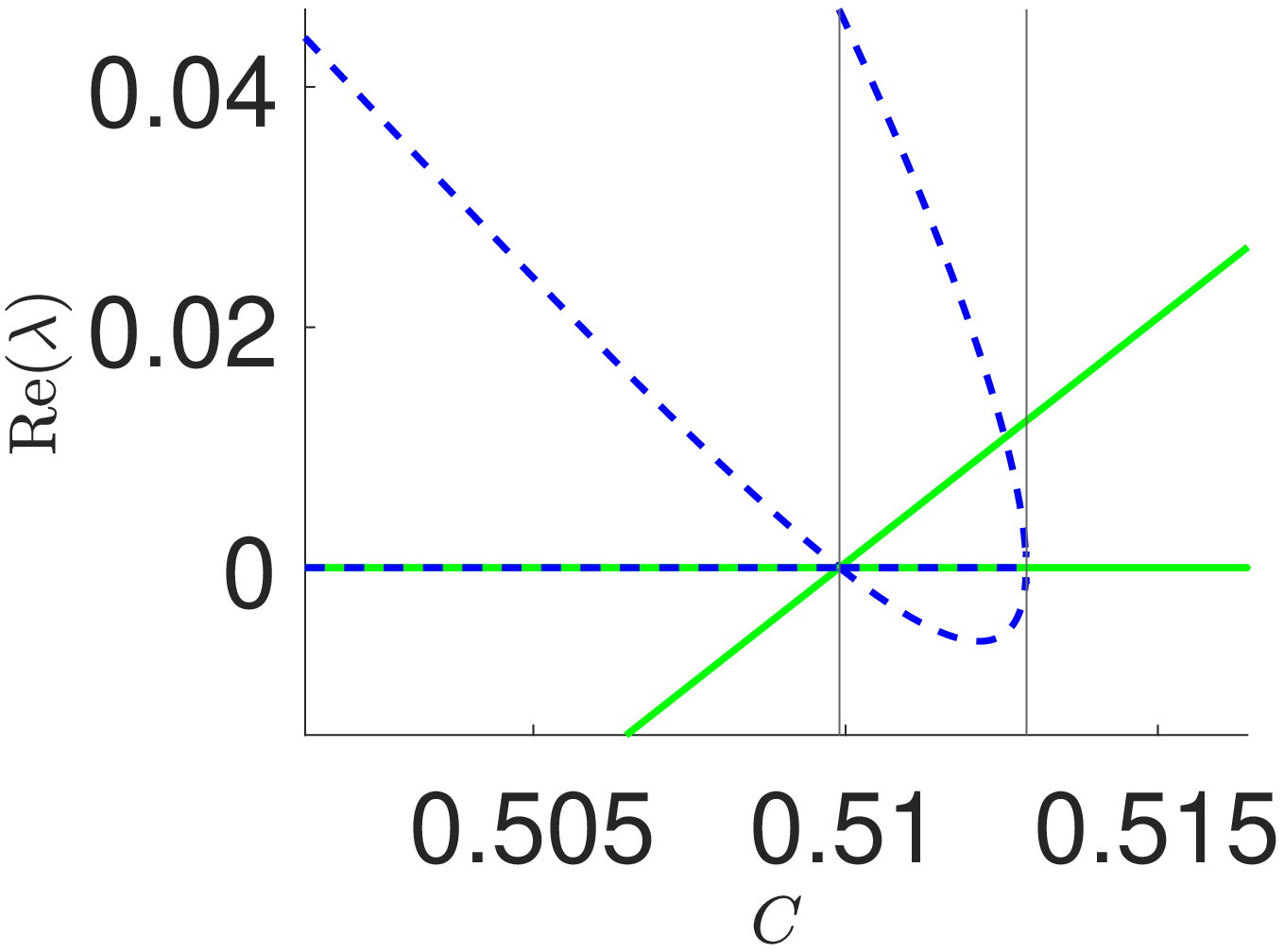}\quad
                \includegraphics[width=.4\textwidth]{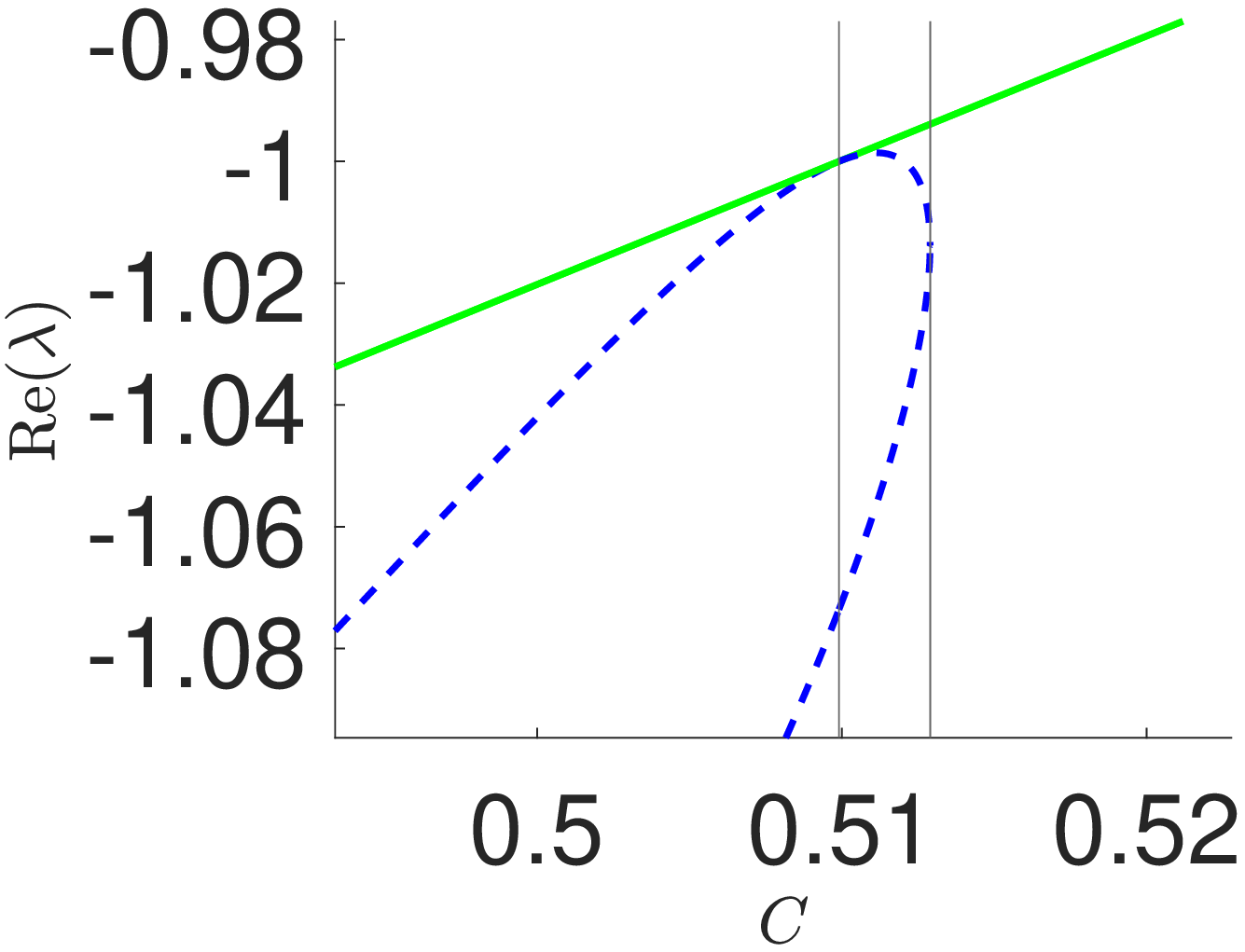}\quad
		\caption{{Eigenvalues of the asymmetric (and lower
                    symmetric)
                    branches as a function of (control)
			parameter $C$. Top left: real part; top right:
                        imaginary part. The second and third row show
                        details of the individual eigenvalues
                      in the vicinity of the bifurcation points
                      $C_{cr}^{(1)}=0.51$ (transcritical) and
                      $C_{cr}^{(2)}=0.5128$ (SN), marked by thin black lines. Once again green (solid)
                      lines are used for the lower symmetric branch,
                      while dashed (blue) ones for the asymmetric
                      branches. See the text for further discussion.}}
		\label{fig:eigs_vs_C}
	\end{minipage}
\end{figure}

  The conversion of these complex pairs into real ones is
  also manifest explicitly in the top panels of the detailed
  Fig.~\ref{fig:eigs_vs_C},
  which constitutes a central set of our numerical findings.
  Indeed, the top right panel shows how the complex pairs collide
  at the above critical points, thereafter splitting into two real
  eigenvalues
  for the respective branches, as shown in the top left panel of
  Fig.~\ref{fig:eigs_vs_C}. The left panel, admittedly, becomes
  rather complicated as we approach the critical points
  $C_{cr}^{(1)}$ and  $C_{cr}^{(2)}$ although a collision with the
  green (lower symmetric) branch is apparent. To that effect,
  we provide further details in the middle and bottom row panels
  of this Figure, where the detail of each of the relevant eigenvalues
  is shown in the vicinity of $C_{cr}^{(1)}$ and  $C_{cr}^{(2)}$,
  i.e., close to the transcritical and the SN bifurcation points, 
  respectively. The ``collision'' of the asymmetric lower and
  symmetric branch is evident in these three panels at $C_{cr}^{(1)}$.
  Especially telling within the right panel of the middle row is
  the exchange of stability between the symmetric (lower) green
  branch and the asymmetric branch. Notice that both branches
  already bear a real eigenvalue (hence are unstable). However,
  the asymmetric branch has a second eigenvalue crossing
  from positive to negative, while the symmetric one goes in 
  the opposite direction, with the two exchanging their stability
  in the aforementioned transcritical bifurcation event. Lastly,
  the asymmetric branch terminates at $C_{cr}^{(2)}$ through
  a SN bifurcation featured in the middle right panel through
  two eigenvalues colliding at $0$. We believe that this description
  offers a comprehensive understanding of the bifurcation
  phenomenology
  present in the system.
  
\subsection{Dynamics}
\begin{figure}[!ht]
	\centering
        \centering
        \includegraphics[scale=0.55]{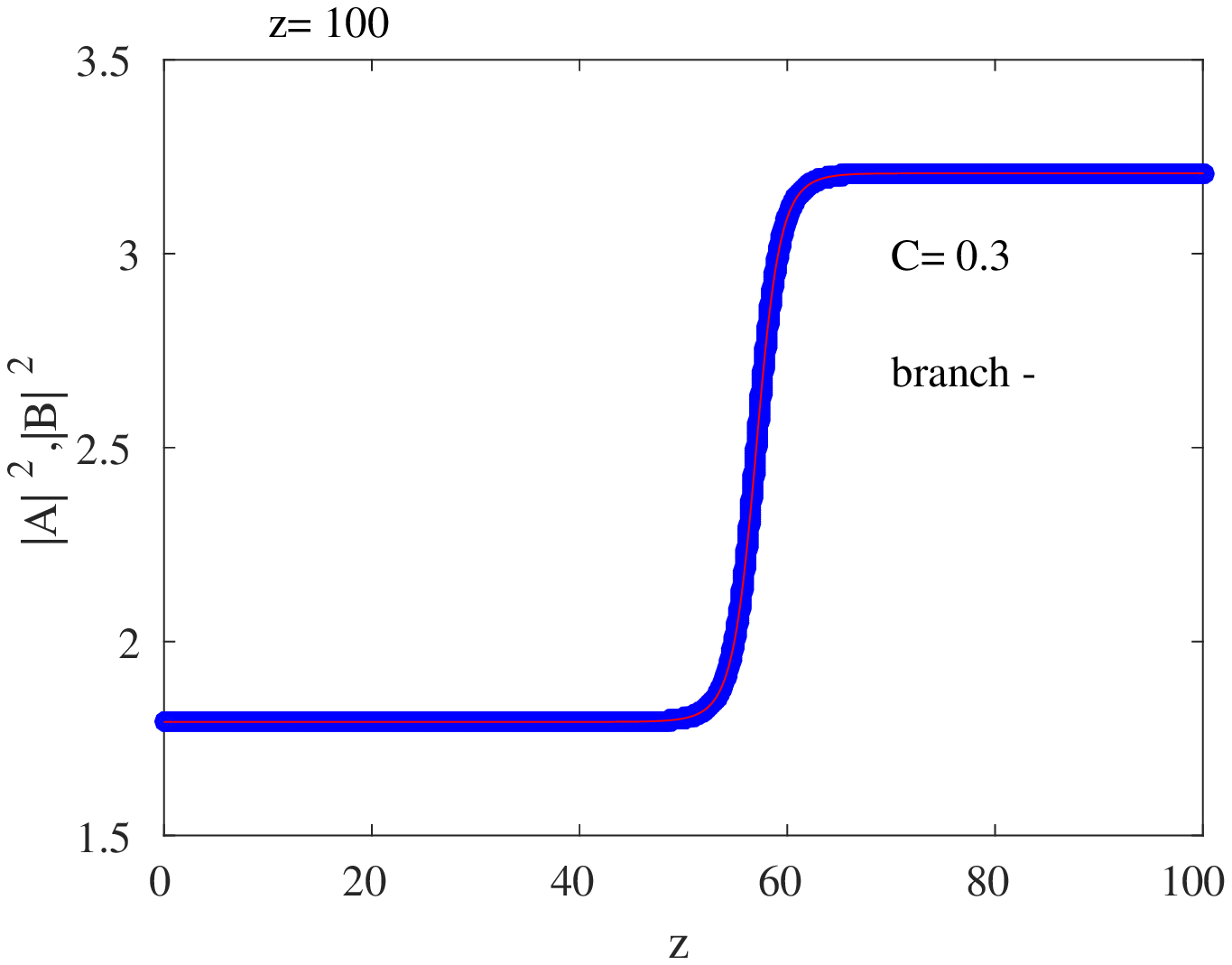}		
		\includegraphics[scale=0.55]{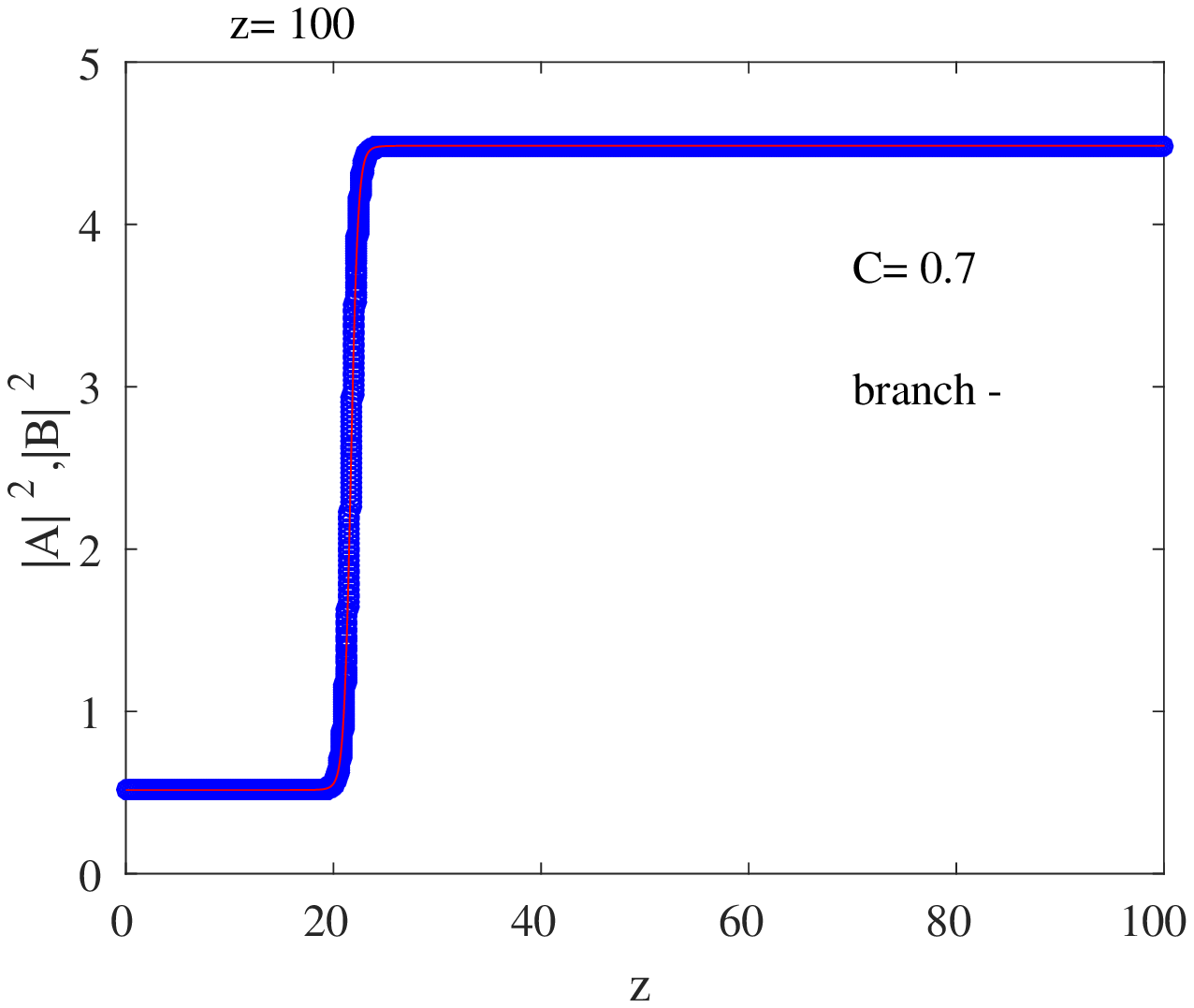}
		\includegraphics[scale=0.55]{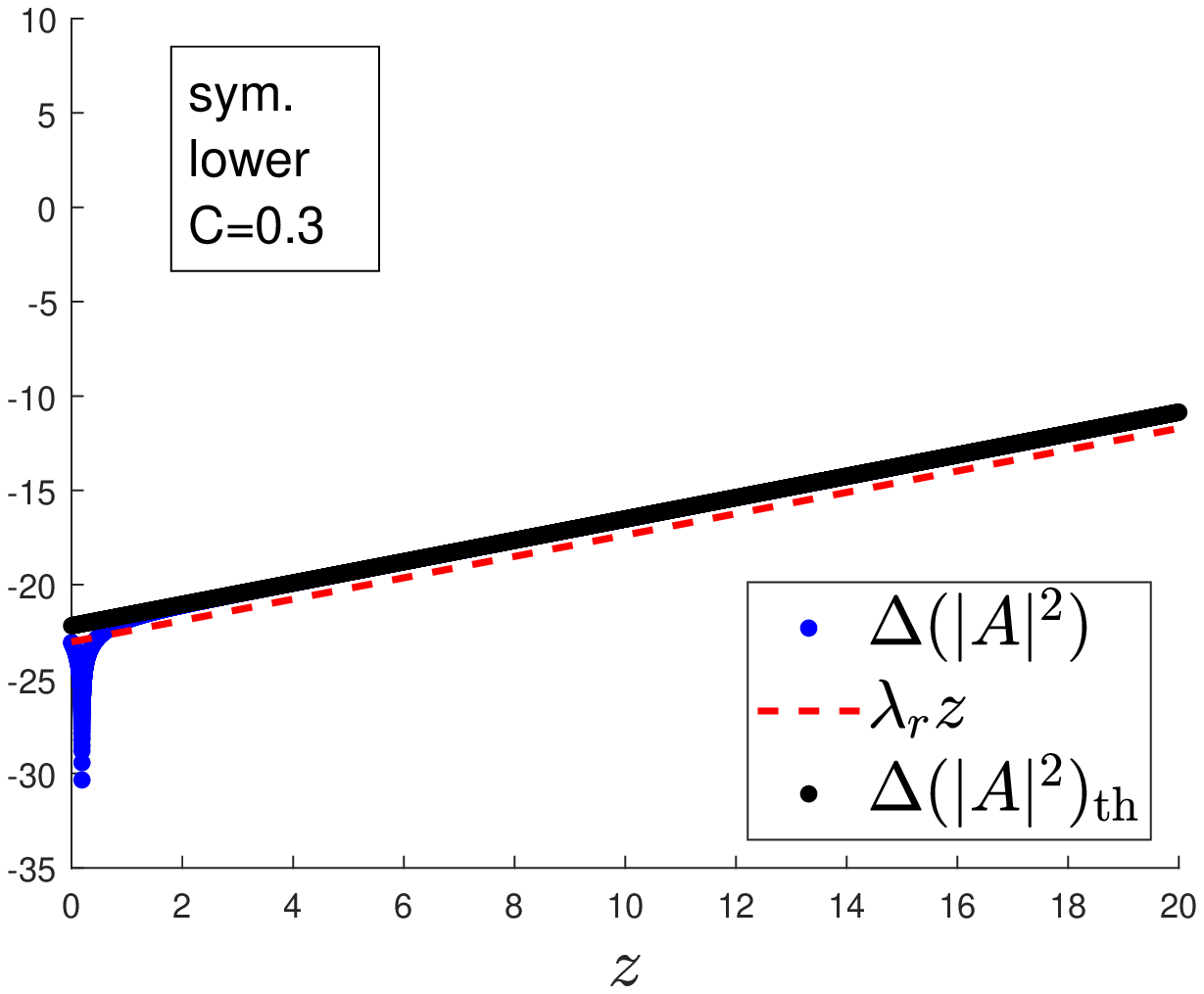}
		\includegraphics[scale=0.55]{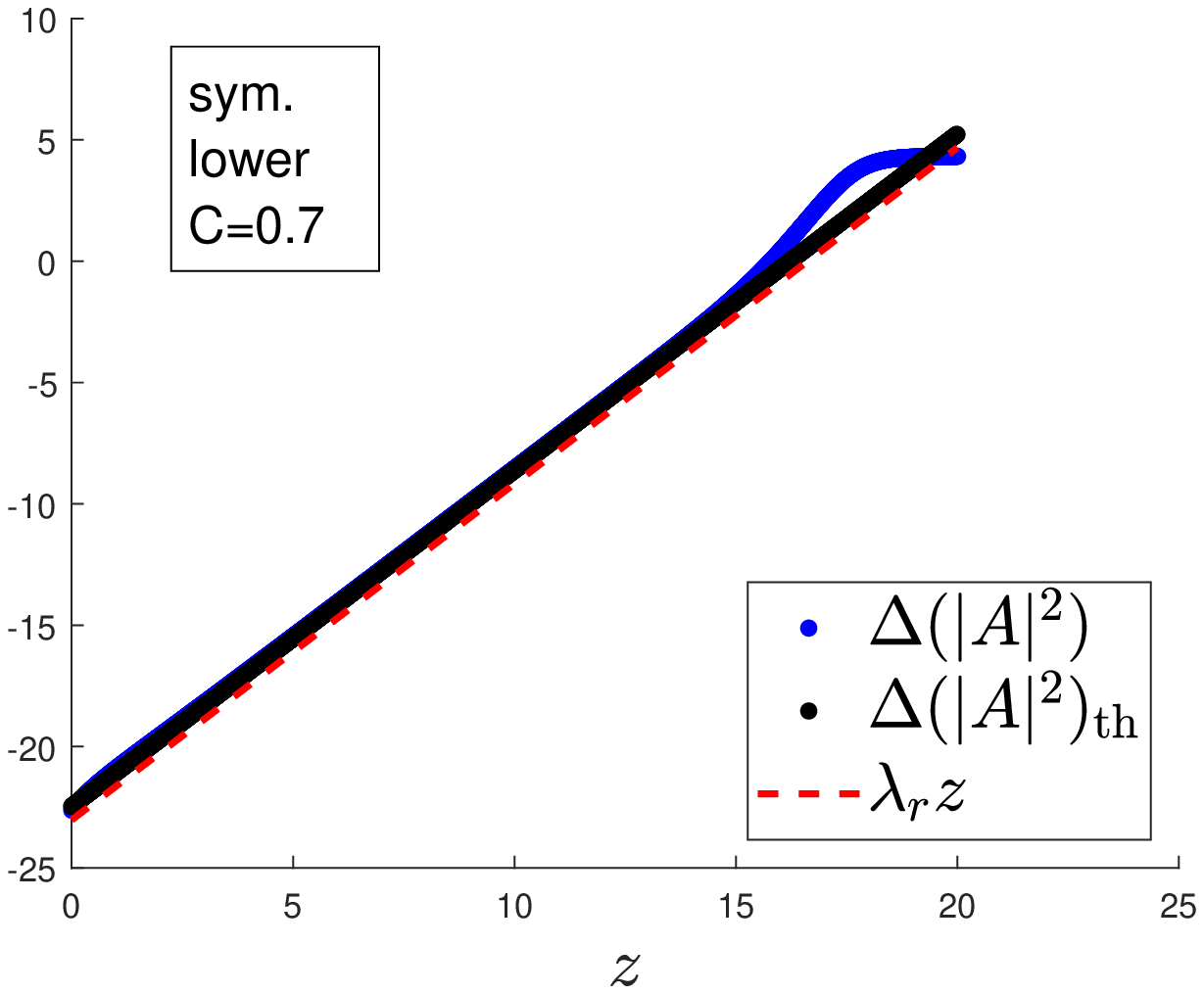}
	\caption{Evolution of symmetric steady state solutions from the
          lower branch in linear (top) and semilog (bottom) scale. The left
          panels
          are for $C=0.30$ and right ones for $C=0.70$.
          Other parameter values as in previous figures. It is clear
          (from the top panels) that the evolution tends to the stable
          symmetric (upper branch) structures.
        } 
	\label{fig:logdyn-s3}
\end{figure}

Guided by the stability results we evolved initial conditions of both
branches and both types of solutions for $C$ values that should
illustrate some of the principal features of the stability diagrams picture.
In the case of the symmetric, upper branch we verified that initiating
the
dynamics along this branch yields a perfectly stable dynamical
evolution, even upon perturbation of the branch (results not shown
for brevity). 
On the other hand, the initial conditions belonging to the lower symmetric
branch evolve towards the upper branch, as may be expected, given that
for both $C=0.3$ and $C=0.7$ it has an eigenvalue with a positive real
part and the only stable solution of the system
is the upper symmetric one. This is shown in the top panels of Fig.~\ref{fig:logdyn-s3}.

To illustrate the relevant instability more clearly (and its
connection
with the spectral picture that we have previously obtained),  we
perturb the initial condition (steady state) with the eigenvector
corresponding to the
eigenvalue with the largest real part, in order to accelerate the
decay and to
check
if the evolution corresponds indeed to the growth at a rate
associated with the real part of the eigenvalue,
$\lambda_{r,\text{max}}$ (the maximal positive real eigenvalue).
We present these results for the lower branch, both for $C=0.30$ and
for $C=0.70$
in the lower panel of Fig.~\ref{fig:logdyn-s3}. We plot the semilog of
the variation in power
relative to the steady state (subscript ss) solution,
$\log\big(|A(z)|^2-|A_{ss}|^2\big)$.
As evidenced in the figure, the growth slope matches very accurately 
the real part of the (most unstable) eigenvalue, confirming the
results of our spectral analysis.

\begin{figure}[!ht]
		\includegraphics[scale=0.5]{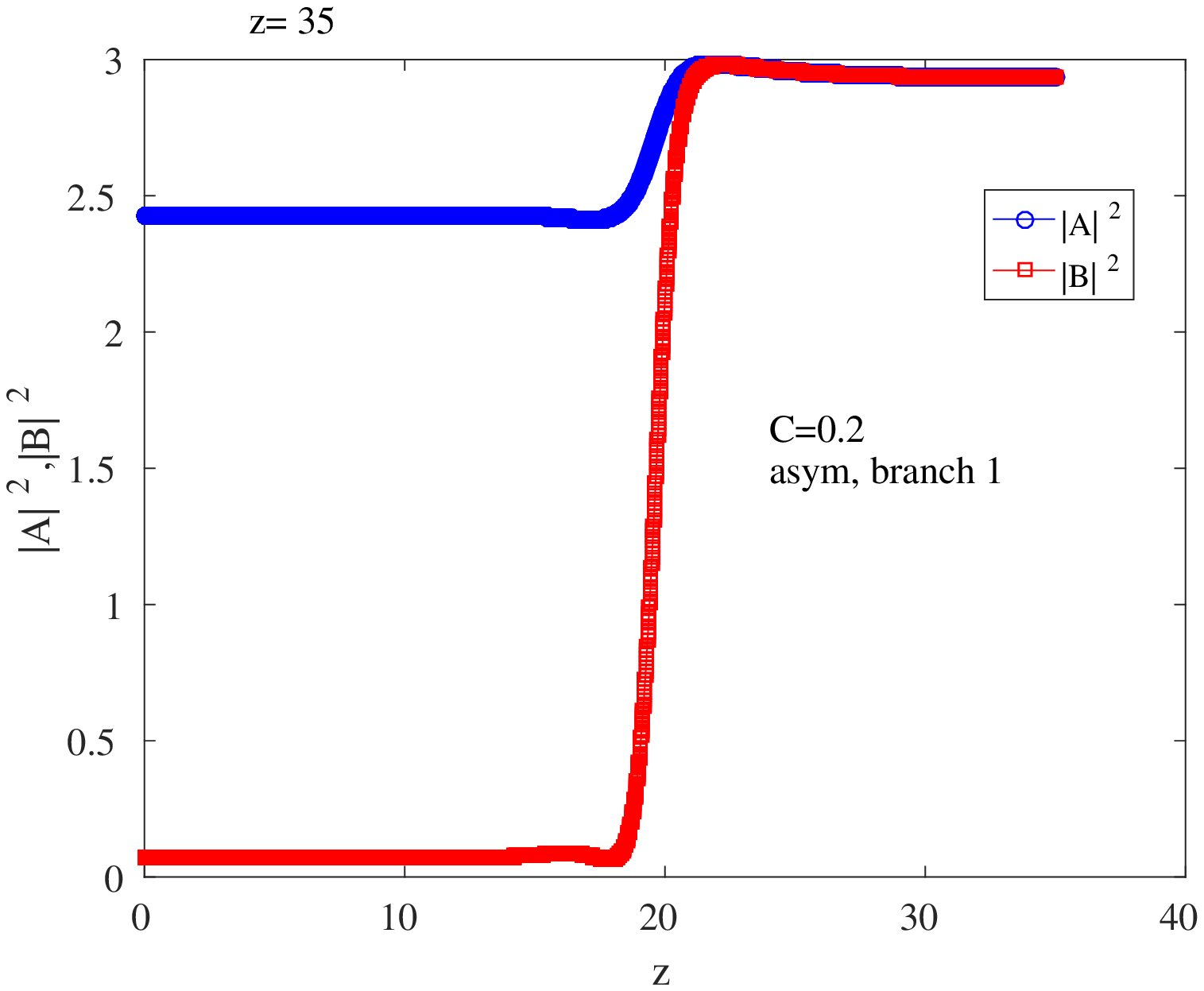}
		\includegraphics[scale=0.5]{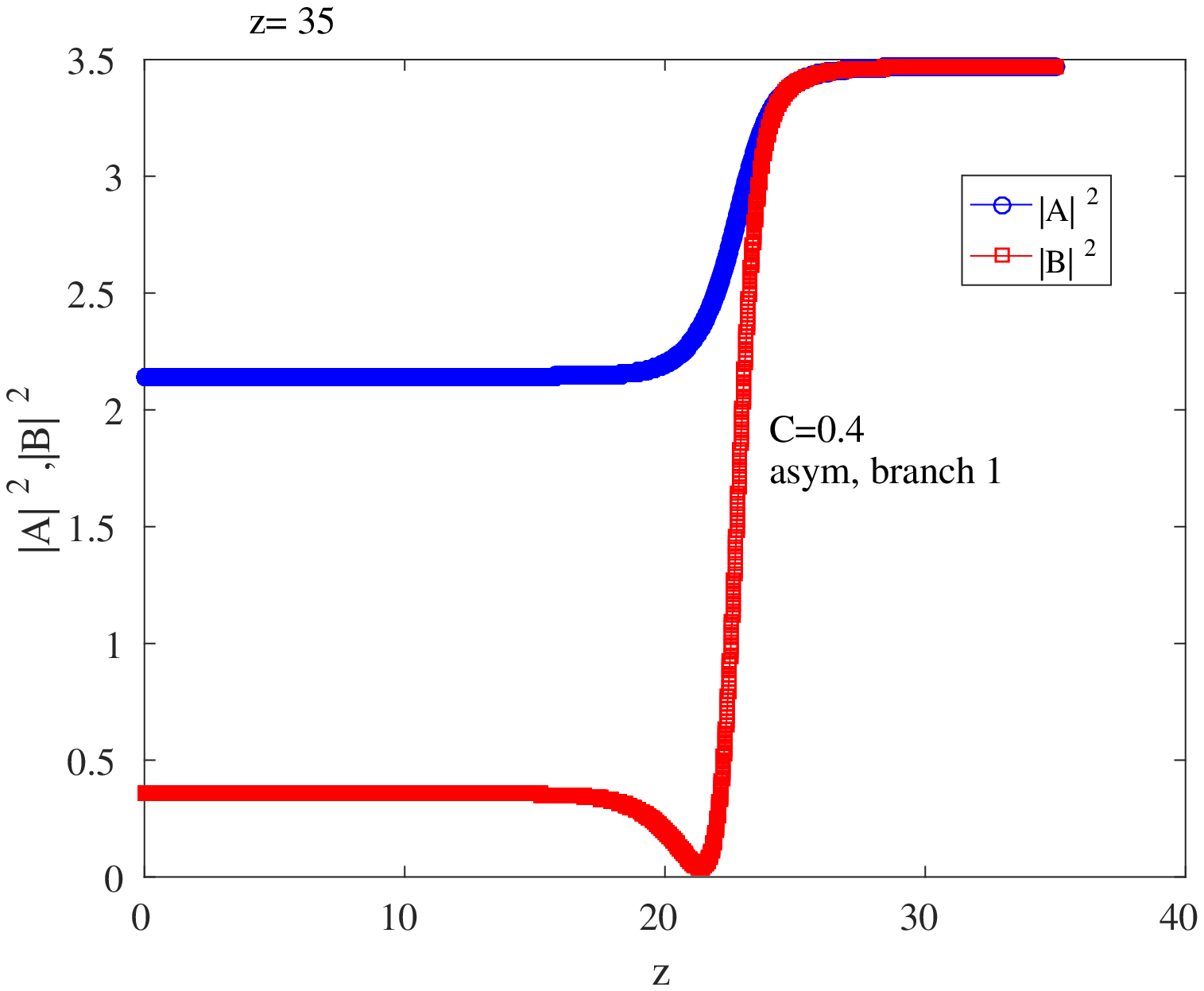}
		\includegraphics[scale=0.5]{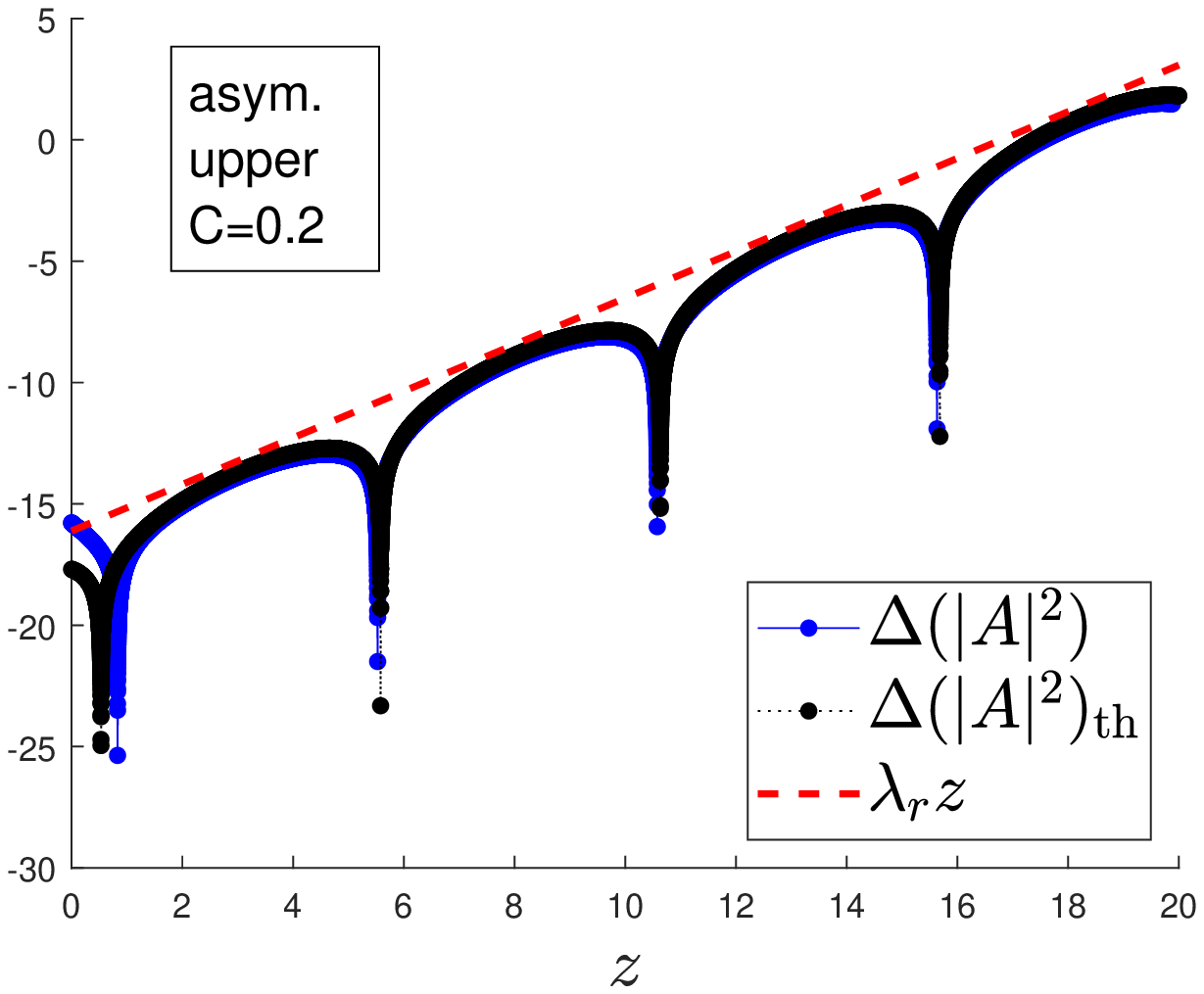}
		\includegraphics[scale=0.5]{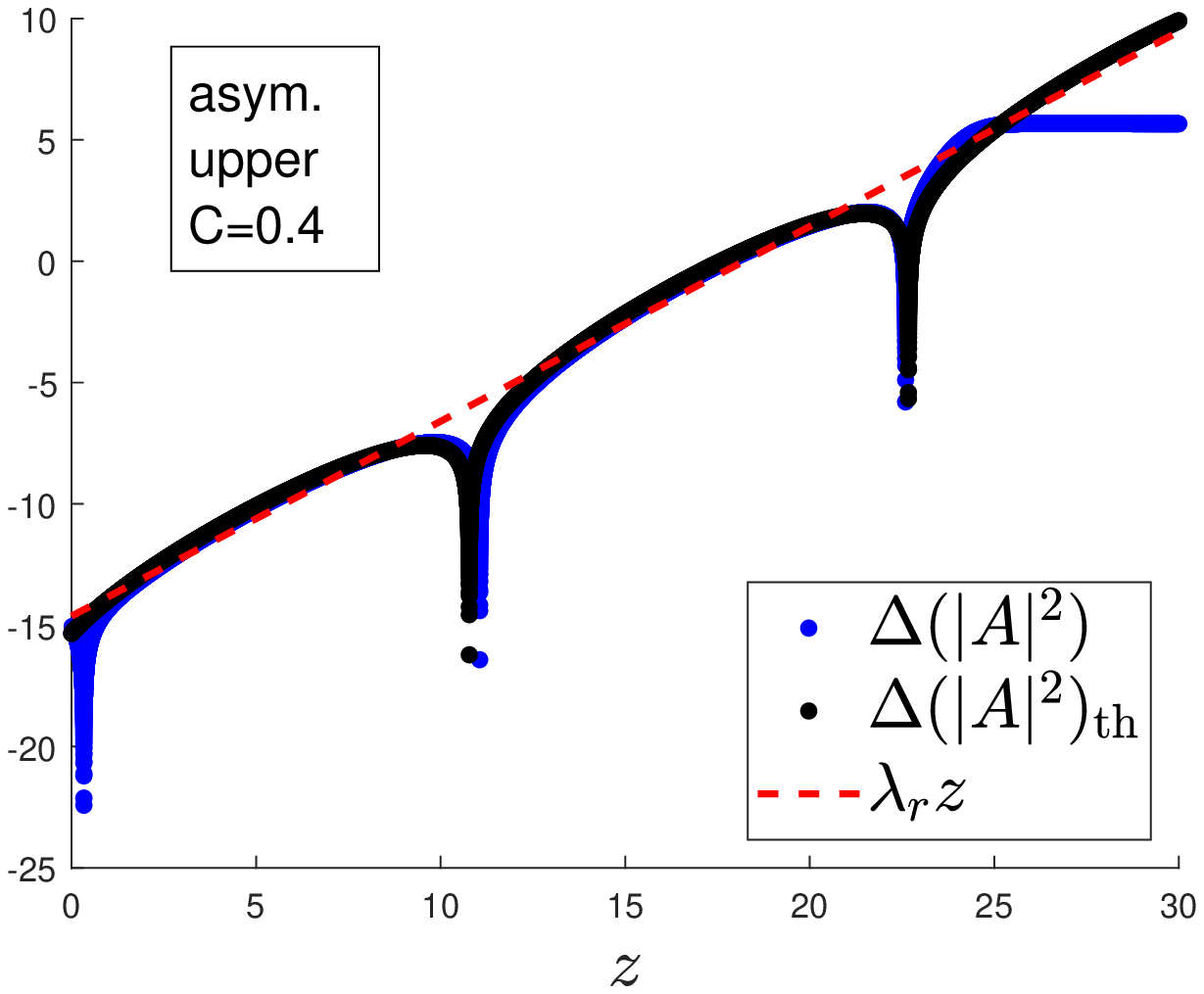}
	\caption{Evolution of steady state solutions from the asymmetric
          upper branch (a) $C=0.2$ (b) $C=0.4$. The amplitudes of the
          top
        panel evolve towards the symmetric values of the upper
        symmetric
        (stable) branch in the top panels. The bottom panels show
        the growth process in semilog scale
        corroborating not only the real part
      involving
    the growth (dashed red line), but also the imaginary part
    associated
  with the oscillation (cf. the theoretical curve in black vs. the
  numerical
results in blue). } 
	\label{fig:dyn-s4}
\end{figure}

\begin{figure}[!ht]
	\centering
		\includegraphics[scale=0.45]{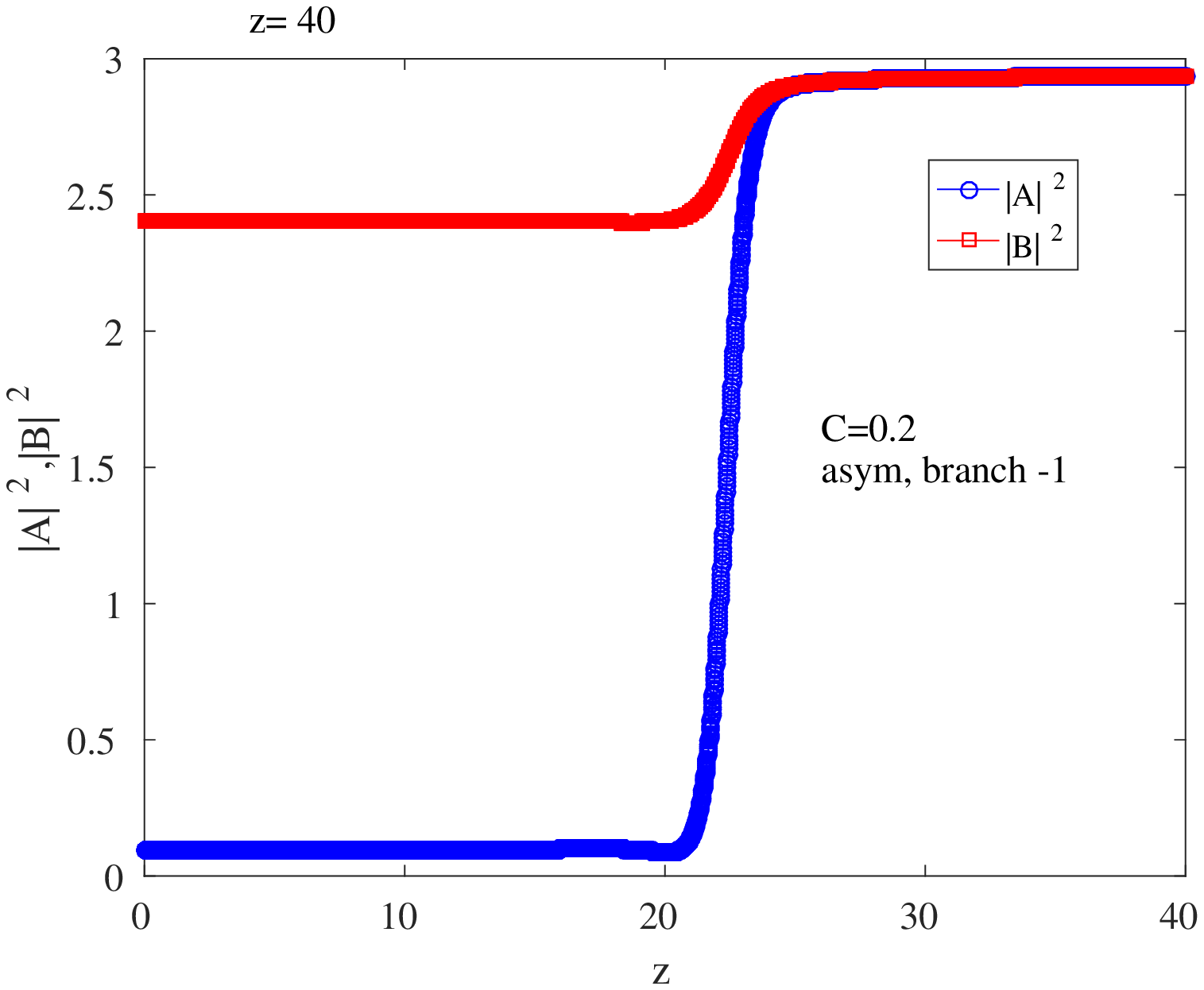}
		\includegraphics[scale=0.45]{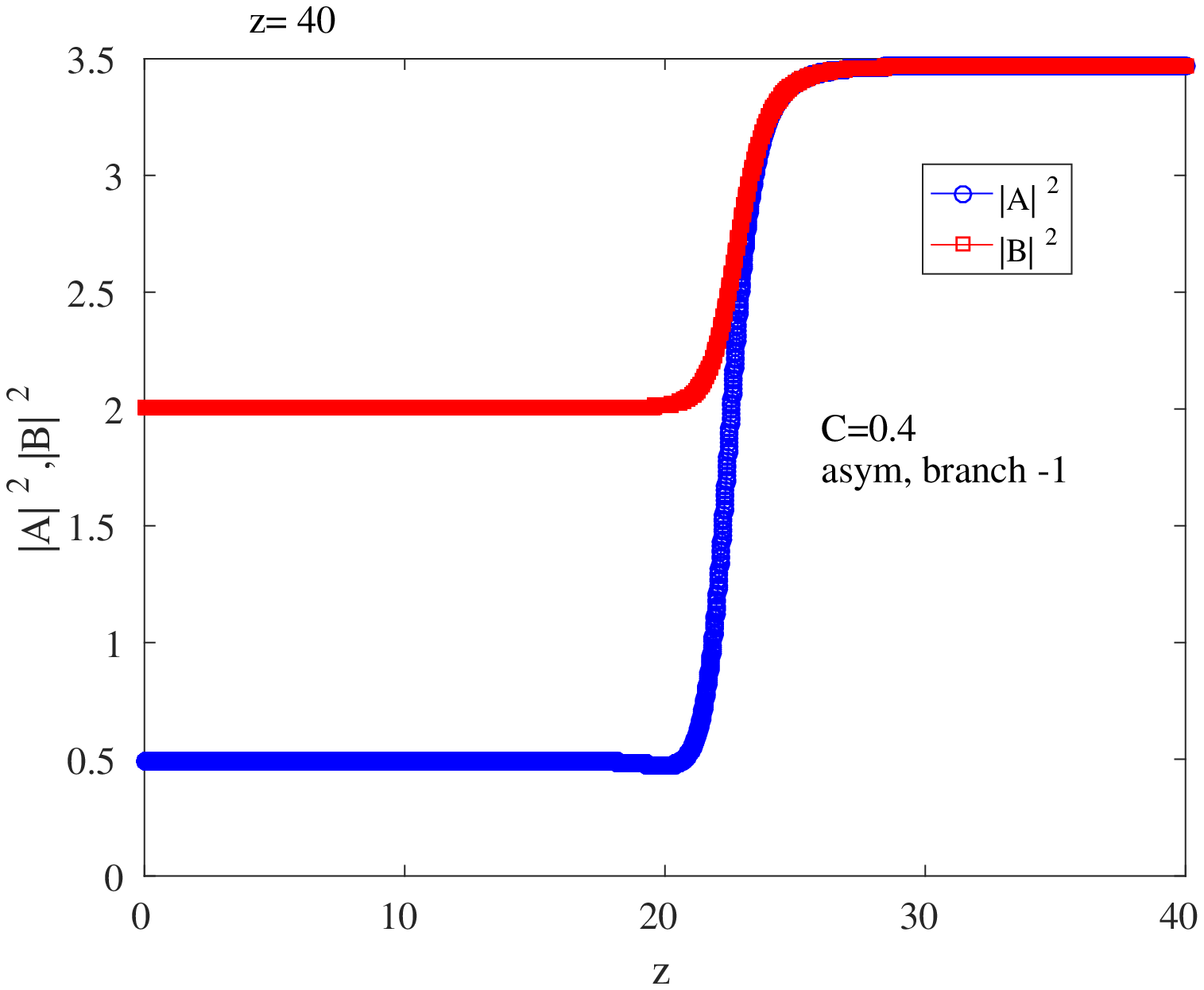}
        \includegraphics[scale=0.45]{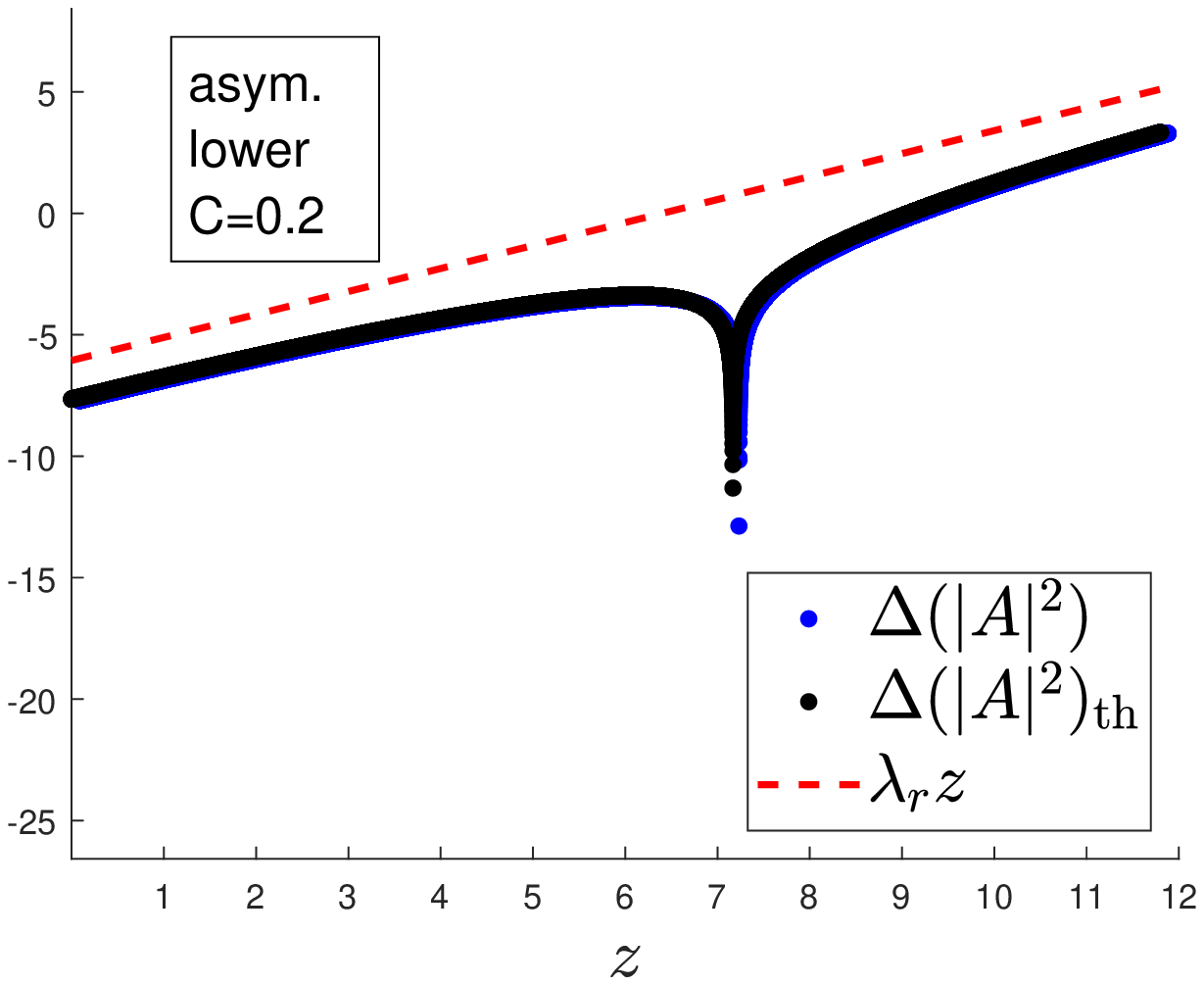}
		\includegraphics[scale=0.45]{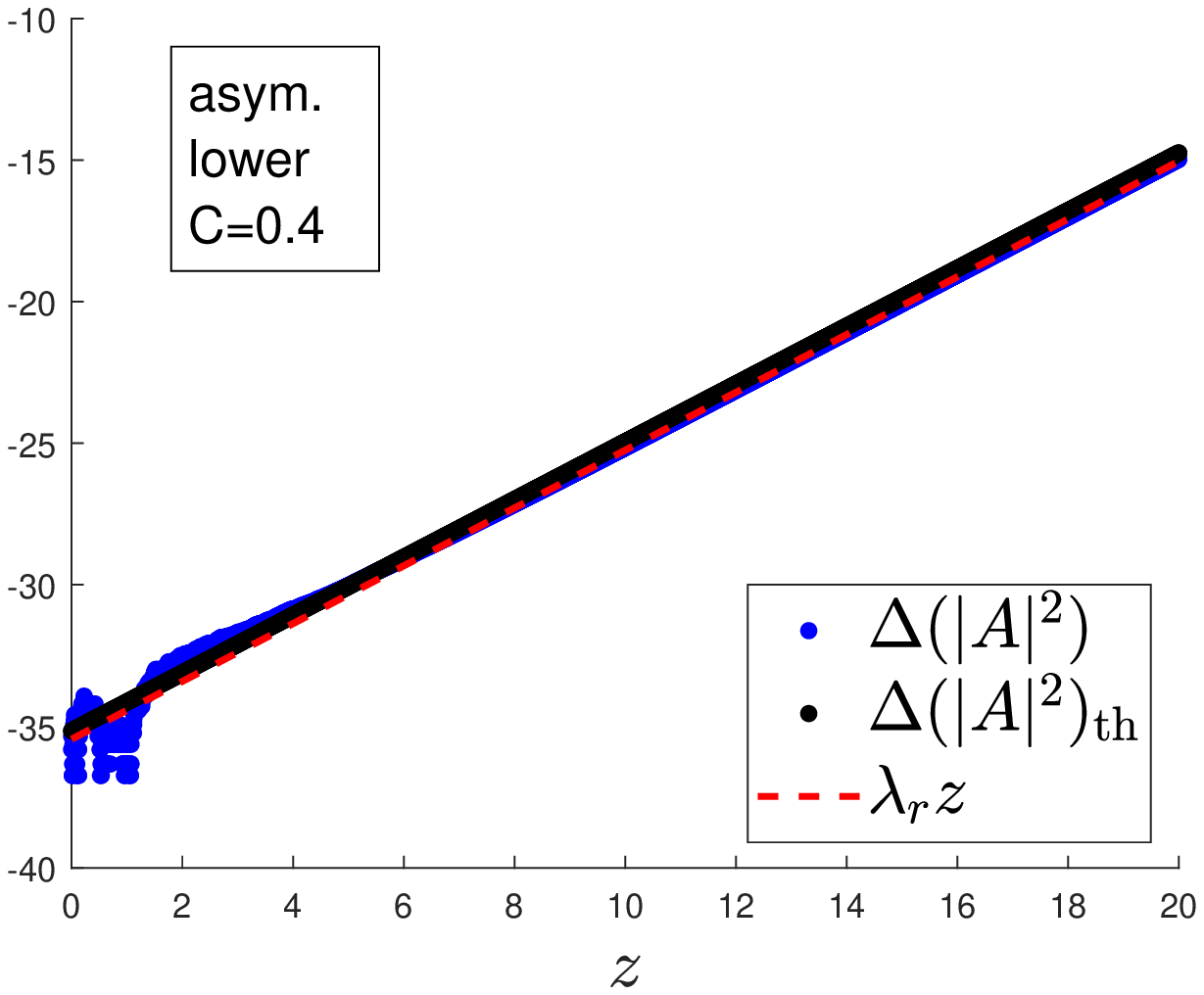}
	\caption{Evolution of steady state solutions from the asymmetric
          lower branch (a) $C=0.2$ (b) $C=0.4$. See text for other
          parameter values.
        The figure is similar to Fig.~\ref{fig:dyn-s4}, however for
        this
        branch the case of $C=0.2$ has a complex pair, while that
      of $C=0.4$ possesses only real eigenvalues; cf. Fig.~\ref{fig:asymua}.} 
	\label{fig:dyn-s5}
\end{figure}

Now let us look at the dynamics of the asymmetric solutions, both for
the 
upper and lower branch, illustrated in
Figs~\ref{fig:dyn-s4}-\ref{fig:dyn-s5}, respectively.
As predicted by linear stability, in both cases it is perceivable that
the initial state evolves towards a symmetric state, and from the
final amplitude it is the upper symmetric state, i.e., the only
linearly stable configuration available in the system.
This is shown in the linear scale plots of the top panels. On the
other hand,
we also present the semilog plots of the evolution of the departure
from initial steady state. In this case we also represent the 
theoretical curve for the prediction for the evolution of the
perturbation along the eigenvector with largest real part;
this curve is denoted $\Delta(|A|^2)_{{\rm th}}$. 
Similar to the (lower branch) symmetric case, the plot of $
\Delta(|A|^2) = \log(|A(z)|^2-|A_{ss}|^2)$  in
Figs.~\ref{fig:dyn-s4}-\ref{fig:dyn-s5} shows a relation to the
eigenvalue,
as the slope of the tangent to the curve.
However, in this case, the relevant eigenvalues are complex, hence
there is not only a growth associated with the real part of the
eigenvalues,
but also an oscillation associated with the imaginary part of the
pertinent eigenvalue. This oscillation is clearly evident in the
bottom panel of both figures, and it indeed matches the expected one
on the basis of the imaginary part of the eigenvalue. This
definitively
corroborates the spectral results of our stability analysis.
Recall, however, from Fig.~\ref{fig:asymua} that the lower
asymmetric branch has a purely real instability for $C=0.4$
(while it has a complex pair for $C=0.2$). This is also
corroborated by the results of Fig.~\ref{fig:dyn-s5},
by comparing the exponential growth of the former
case (right panels) with the oscillatory one of the latter case
(left panels).

\section{Conclusions \& Future Work}
\label{sec:conclusion}

In the present work, we have explored a nonlinear variant of the
anti-$\PT$ symmetric dimer problem. The linear version of this
setup has already been explored in a variety of settings, including 
optical waveguides~\cite{GeTur13,YaLiYo17,lige},
coupled electrical circuit resonators~\cite{Choi18} and atomic
vapour cells~\cite{PengXiao}. Some of these works have already
proposed variants of the relevant settings that would involve
nonlinearity~\cite{PengXiao}, while for others we argued about
the fact that nonlinearity inclusion would be natural on the basis
of the nature of the response of such systems at larger amplitudes.
We have explored the most prototypical nonlinear dimer setting
and were able, given the few-degrees-of-freedom nature of the
setting, to obtain solutions analytically for the stationary states
of the system. We found, in particular, two symmetric solutions
arising via a saddle-node bifurcation and also identified two
asymmetric solutions which are involved in a
transcritical bifurcation with the lower symmetric branch, as well
as in a saddle-node bifurcation leading to the termination of
the asymmetric solutions.  Out of these four solution branches,
only one was found to be spectrally stable and indeed was
identified as a generic attractor of the dynamics of the system,
even when starting from the unstable symmetric or asymmetric
solutions. Our spectral analysis was straightforwardly corroborated 
via direct numerical simulations of the evolution dynamics which
showed the growth along the predicted unstable eigendirections
of unstable stationary states with the appropriate rates, and
the eventual approach to the sole dynamical attractor of
this system, namely the stable (upper) symmetric branch.

Naturally, these results pave the way for numerous further
studies of anti-$\PT$ symmetric systems along a similar
vein to what was done in the $\PT$-symmetric case~\cite{RevPT,KZY,JY}.
In particular, one can examine so-called anti-$\PT$ symmetric
oligomers ($\PT$-symmetric ones were explored, e.g.
in~\cite{lik,LiKeMaGu2012,KoZe2012}), as well as
lattices of such elements (again, corresponding $\PT$-symmetric
explorations could be found in~\cite{tuyg1,tuyg2}).
This can be done both in one- but also in higher dimensions.
Furthermore, here, we have concerned ourselves with cubic Kerr-type
nonlinearities, yet some of the above settings seem to be well-suited
for different types of nonlinear terms, including four-wave-mixing
ones~\cite{PengXiao}, with the latter being another topic worthwhile
of further study. Such considerations are currently in progress and
will be reported in future publications.

\section*{Conflict of Interest Statement}

The authors declare that the research was conducted in the absence of any commercial or financial relationships that could be construed as a potential conflict of interest.

\section*{Author Contributions}

ASR: Data curation, Investigation, Software, Validation,
Visualization, Writing - original draft;
RMR: Data curation, Investigation, Software, Validation,
Visualization, Writing - original draft;
VVK: Conceptualization, Methodology, Investigation, Writing - review
\& editing;
AS: Conceptualization, Methodology, Investigation, Writing - review \& editing.
PGK: Conceptualization, Methodology, Investigation, Validation, Supervision, Writing - original draft.


\section*{Funding}
A.S.R acknowledges financial support from FCT-Portugal through Grant No. UIDB/04650/2020.
This material is based upon work supported
by the US National Science Foundation under Grants
No. PHY-2110030 and DMS-1809074 (PGK).
VVK acknowledges financial support from the Portuguese Foundation for Science and Technology (FCT) under Contract no. UIDB/00618/2020. The work of A.S. at Los Alamos National Laboratory was carried out under the auspices of the U.S. DOE and NNSA under Contract No. DEAC52-06NA25396 and supported by U.S. DOE.




\bibliographystyle{frontiersinHLTH&FPHY} 


\end{document}